\documentclass[useAMS,usenatbib,usegraphicx]{mn2e}

\usepackage{todonotes}
\newcommand{\mybf}{}

\title[Simplified Models of Planetary Formation]{Statistics and
  Universality in Simplified Models of Planetary Formation}

\author[C. Hern\'andez-Mena and L. Benet]%
{C. Hern\'andez-Mena$^{1,2}$\thanks{E-mail: cmena@fis.unam.mx}
and
L. Benet$^{1}$\thanks{E-mail: benet@fis.unam.mx}\\ 
$^{1}$ Instituto de Ciencias F\'{\i}sicas, Universidad Nacional Aut\'onoma de 
M\'exico (UNAM), A.P. 48--3, 62210 Cuernavaca, M\'exico\\
$^{2}$ Facultad de Ciencias, Universidad Aut\'onoma del Estado de Morelos (UAEM), 
62209 Cuernavaca, M\'exico}

\begin{document}

\date{\today}


\maketitle

\label{firstpage}

\begin{abstract}
  In this paper, we modify Laskar's simplified model of planetary
  evolution and accretion to account for the full conservation of the total angular
  momentum of the system, and extend it to incorporate an accretion
  probability that depends on the mass and relative velocity of the
  colliding particles. We present statistical results for the mass and
  eccentricity of the planets formed, in terms of their semi-major
  axes, for a large number of realizations of different versions of
  the model. In particular, we find that by combining the
  mass-dependent accretion probability and the velocity-selection
  mechanism, the planets formed display a systematic occurrence at
  specific locations. By introducing properly scaled variables, our
  results are universal with respect to the total angular momentum of
  the system, the mass of the planetesimal disc, and the mass of the
  central star.
\end{abstract}


\section{Introduction}

\begin{quotation}
\begin{tabular}{r}
Essentially, all models are wrong, but some are useful.\\
\citet{BoxDraper}
\end{tabular}
\end{quotation}

Recent advances in the understanding of the final stages of the
formation of the Solar System within the Nice model, have achieved,
among others, reproducing the architecture of the outer part to an
unprecedented level~\citep{Tsiganis2005,Morbidelli2007,Morbidelli2009}. 
In the `first' Nice model~\citep{Tsiganis2005}, the
initial conditions are defined such that Jupiter is close to its
actual position, Saturn is placed just below the semi-major axis
corresponding to the 1:2 mean-motion resonance with Jupiter, and the
icy giants are placed in the intervals 11--13 AU (Uranus) and 13.5--17
AU (Neptune); the planetary masses are the current ones. In addition,
the Nice model assumes a 30--50 M$_{\oplus}$ planetesimal disc placed
beyond the initial orbit of Neptune which ends between 30--35 AU. The
planetesimal disc transfers angular momentum to the planets making
them migrate~\citep {FernandezIp84,Malhotra95}, allowing Jupiter and
Saturn to cross their mutual 1:2 mean-motion resonance. This event
increases their eccentricities and induces secular resonances on the
icy giants. There is a period of crossing orbits which eventually
scatters the icy giants outwards, which enter the planetesimal
disc~\citep{Levinson2003}. The icy giants eject inwards some
planetesimals to Jupiter and Saturn, which makes them migrate even
further away. The planets finish their migration once the planetesimal
disc is completely depleted. The final configuration of the first Nice model
is close to the observed values for both the semi-major axes and the mean
eccentricities of the planets~\citep{Tsiganis2005}. Recently, \citet{Morbidelli2009} 
extended their original results showing that encounters of Saturn with an ice 
giant lead to the correct secular evolution for the eccentricities of Jupiter and 
Saturn, and not the passage through the 1:2 mean-motion resonance, as it 
was originally proposed.

An important open issue in the Nice model is related to explain the
initial conditions assumed in the model, from earlier stages of the
formation. This question has been recently investigated, assuming an
earlier multi-resonant configuration for the already formed giant
planets, besides pushing the planets into such configuration, by
introducing additionally semi-major axis and eccentricity
decay~\citep{Batygin10}. Yet, the issue remains to explain those
initial conditions.

The aim of this paper is to study the statistics of the final
configurations of planetary systems for some models of planetary
accretion and evolution, which are based on a model originally
introduced by~{\mybf\citet{Laskar2000}}. The final orbital parameters we
obtain could be used as the initial conditions for detailed
calculations on the planetary evolution, e.g., aimed to describe the
architecture of the planetary systems formed in the spirit of the Nice
model, or to address the dynamical stability of these systems in
statistical terms. Yet, it is not our purpose to obtain an accurate
comparison with the initial conditions of the Nice model, or with the
architecture of the Solar System. Our aim is to understand the
consequences of incorporating in a simple form some important physical
effects on the accretion processes. The variations of Laskar's model 
included account for the systematic conservation of the total angular 
momentum of the system, and incorporate mass- and velocity-selection 
mechanisms that tune the accretion probability; 
{\mybf note that Laskar's model actually conserves the angular 
momentum, but not the methodological procedure used to achieve 
a fixed final angular momentum deficit (AMD)}. These aspects were not
considered in Laskar's model, though they constitute important
building blocks of the current planetary formation
theories~\citep{Safronov,Pollack96,Armitage:2007arXiv}. By comparing
the outcome of the mechanisms that we include, we are able to relate
them, in a statistical sense, to certain effects manifested by the
final configurations. This is done directly by correlating the masses
and semi-major axes of the planets formed in a large number {\mybf of}
realizations of the models. Our results indicate a power-law behaviour
of the mass in terms of the semi-major axis, whose exponent depends on
the different models, and on the initial linear mass-density
distribution assumed. In particular, we show that including both mass-
and velocity-selection mechanisms leads to the systematic appearance
of planets in certain specific locations. Moreover, by introducing
properly-scaled variables, we show that these results are universal
with respect to the total angular momentum of the system, the mass of
the planetesimal disc and the mass of the central star.

The paper is organised as follows: In Section~\ref{LaskarModel} we
describe Laskar's simplified model of planetary evolution and
accretion in detail, clarify its predictions, and outline a short
critique aimed to motivate the variations that we incorporate in the
model. Specific details of the variations we implement and the
corresponding numerical results are described in Section~\ref
{variations}. In Section~\ref{univ} we address how variations of the
total angular momentum of the system, the mass of the planetesimal
disc, and the mass of the central star affect our results.  Finally, in
Section~\ref{concl} we summarise our results and conclusions.

\section{A simplified model of planetary formation and evolution}
\label{LaskarModel}

\subsection{The model}

Laskar's simplified model of planetary formation and evolution
{\mybf\citep{Laskar2000,LaskarPrep}} considers a
system which consists of one large central body of mass $M_0$ and a large
number of particles of small mass $m_i$, $i=1\dots N$ ($N\gg 1$), that
interact under their mutual gravitational attraction.  The
Hamiltonian of this $(N+1)$--body problem, in a heliocentric frame, can
be written as $H=H_0+H_1$, where $H_0$ and $H_1$ are given by%
\begin{eqnarray} \label{eq:hamiltonian0}
  H_0 & = & \sum_{i=1}^{N} \Big[ \frac{{\mathbf p}_i^2}{2 m_i} - 
  \frac{{\cal G} M_0 m_i}{r_{i}} \Big],  \\
\label{eq:hamiltonian1}
  H_1 & = &  \frac{1}{2 M_0} {\mathbf P}^2_0 - \sum_{1\le i<j}^{N} 
  \frac{{\cal G} m_i m_j}{r_{ij}}. 
\end{eqnarray}
Here, ${\cal G}$ is the gravitational constant, $r_i$ is the mutual
distance between the $i$-th planetesimal and the central star,
$r_{ij}$ is the distance between the $i$-th and $j$-th planetesimals,
${\mathbf P}_0=-\sum_{i=1}^{N} {\mathbf p}_i$ and ${\mathbf p}_i$ are the
canonical momenta associated with the star and the planetesimals,
respectively. Clearly, $H_0$ corresponds to $N$ independent two-body
Kepler problems among each planetesimal and the star, and $H_1$ is the
perturbing term, which includes the mutual gravitational interaction
among the planetesimals and the indirect term. 
{\mybf Laskar's model is formulated for the spatial problem; the numerical 
simulations are illustrated for the planar case. From now on, we focus
on the latter case considering coplanar orbits}.

The dynamics of $H_0$ is completely integrable: For each planetesimal
the energy $E_i=-\mu m_i/(2 a_i)$ and the barycentric angular momentum
$l_i= m_i [\mu a_i(1-e_i^2)]^{1/2}$ ($\mu = {\cal G} M_0$) are
conserved quantities.  This implies the conservation (for $H_0$) of
the semi-major axis $a_i$ and eccentricity $e_i$ of the elliptic
orbits of the planetesimals. The non-integrable character of $H$ is
manifested in the long-term evolution and is due to the perturbing
terms of $H_1$. To model these non-integrable effects,
{\mybf\citet{Laskar2000}} considers the secular system, where the equations
of motion are averaged over the mean longitudes. This simplification
is tantamount to excluding effects related with mean-motion
resonances.

Firstly, considering the case without collisions, the averaged system
conserves the energy of each planetesimal, but exhibits a slow chaotic
diffusion of the individual eccentricities $e_i$. This chaotic
diffusion is constrained by the conservation of the total angular
momentum of the system%
\begin{equation}
  L_{\rm tot} = \sum_{i=1}^{N} l_i= \sum_{i=1}^{N} m_i [\mu a_i]^{1/2}
  (1-e_i^2)^{1/2}.
\end{equation}
Including the conservation of the energy, this leads to the
conservation of the total angular momentum deficit (AMD) of the
system. The latter is expressed as
\begin{equation}
\label{eq:AMDtot}
  C_{\rm tot} = \sum_{i=1}^{N} C_i= \sum_{i=1}^{N} m_i [\mu a_i]^{1/2}
  [1-(1-e_i^2)^{1/2}].
\end{equation}
Physically, the total AMD of a planetary system is a measure of the
circularity of the planetary orbits, since for small eccentricities
$C_i\sim m_i [\mu a_i]^{1/2} e_i^2$.  Conversely, large values of 
$C_{\rm tot}$ indicate the possibility of planetary collisions.

Pair-wise collisions in this simplified model are assumed to be totally
inelastic and lead to accretion. Assuming that mass and linear
momentum are conserved in the collision, $m_{i\oplus j} =m_i+m_j$
and ${\mathbf p}_{i\oplus j}={\mathbf p}_i + {\mathbf p}_j$, the conservation of
the angular momentum follows. From these conservation laws, the
orbital elements of the accreted particle ${i\oplus j}$ can be
calculated: First, the semi-major axis of the accreted particle is
obtained from its energy, which can be expressed as $E_{i\oplus j} =
E_i + E_j + \delta E_{i,j}$. The change in the energy is given by
\begin{equation}
\label{eq:energy}
  \delta E_{i,j} = - \frac{1}{2} \frac{m_i m_j}{m_i+m_j} ({\mathbf v}_i-{\mathbf v}_j)^2 .
\end{equation}
Then, using the conservation of angular momentum of the colliding
particles yields the eccentricity of the accreted particle.

As a crucial final element, {\mybf\citet{LaskarPrep}} demonstrates that
under this accretion scheme, the local AMD of the accreted particles
satisfies
\begin{equation}
\label{eq:AMDacc}
C_{i\oplus j} \le C_i+C_j.
\end{equation}
This fact motivates the so-called AMD stability
criterion~{\mybf\citep{Laskar2000}}: A planetary system is AMD stable if its
total AMD is not sufficient to permit planetary collisions. This
criterion suffices to ensure the long-time stability of the averaged
system, since new collisions are not possible; however, this is not
necessarily true for the complete system $H$, since the latter
includes effects linked with short-period resonances which could still
induce collisions.

Algorithmically, the model is implemented as
follows~{\mybf\citep{Laskar2000}}.  Initially, the semi-major axes are
distributed homogeneously throughout the disc and their associated
masses are fixed by the linear mass distribution $\rho_a(a)$;
likewise, the initial eccentricities are drawn from the distribution
$\rho_e(e)$. The planetesimals are labelled in non-decreasing order
according to their semi-major axis: $a_1 < a_2 < \dots < a_N$, where
$N$ is the initial number of planetesimals. Note that no angular
variables are specified so far. The iterations proceed as follows: One
planetesimal $i$ is chosen at random from the list of planetesimals
together with one of its neighbours, say $i+1$. If the co-focal
ellipses associated with their trajectories display an intersection
--in the purely geometrical sense--, which is expressed as
\begin{equation}
\label{eq:colision}
a_i(1+e_i) \le a_{i+1}(1-e_{i+1}) ,
\end{equation}
then the particles are accreted at one of the intersection
points. Note that the relative orientation of the orbits $\theta$ is a
random variable defined in an appropriate interval, which guarantees
the existence of a geometrical intersection; the necessary condition
Eq.~\ref{eq:colision} thus corresponds to anti-alignment
$(\theta=\pi)$ of the ellipses.  The orbital elements of the accreted
particle are calculated and the list of planetesimals is updated (as
mentioned above, the AMD of the accreted particle is smaller than the
sum of the AMDs of the colliding particles). Between collisions, a
random walk in the space of eccentricities is implemented, which
models the chaotic secular evolution of the
system~\citep{Laskar1989,Laskar1990}, restricted by the conservation
of the total angular momentum. These steps are iterated until the
total AMD of the system is smaller than a critical value $C_{\rm cr}$,
thus ensuring that no planetary collisions are possible.  Clearly, the
final planetary system formed is the result of two competing
processes, a chaotic diffusion due to the secular evolution of the
averaged system that promotes collisions, and the circularisation due
to accretion that inhibits them.

An important simplification of Laskar's model relies on the introduction of 
independent stochastic discrete time-steps or iterations, instead of computing the 
detailed dynamical evolution of the many-body system, which is a complicate and 
time-consuming task. The consequence of this simplification is
the limitation to provide any physically relevant time-scale. Indeed, each iterate 
of the algorithm described above involves the secular chaotic diffusion in the 
space of angular momentum, and the computation of the resulting accreted particle 
from a total inelastic collision of two planetesimals. All details of the actual physical 
processes are abandoned by modelling them as a Brownian motion in the space of 
eccentricities (or equivalently in the space of angular momentum), and by taking a 
random orientation of the elliptical orbits and one of its geometrical intersections.
Additionally, each time-step is computed independently from the previous one, in
the sense that there is no memory. Each of the physical processes involved in
each iteration have different time-scales associated, which depend on the 
semi-major axis, the local mass-density, etc. This implies that two distinct discrete 
time-steps may involve quite different scales of the physical time. Therefore, an 
important limitation of Laskar's simplified model --which is inherited by our variations--, 
is the impossibility to answer questions on the physically relevant time-scale based 
on the simulations.

Despite of the lack of a characteristic time-scale, 
the simplicity of Laskar's model is reflected in the fact that it
permits certain analytical treatment that yields concrete
predictions. These results follow from the minimum possible AMD value
of two colliding particles, that takes place with the outermost
located at perihelion and the innermost at aphelion. After some
algebraic manipulations~{\mybf\citep{LaskarPrep}}, the minimum or
critical AMD $C_{\rm cr}$, which allows for that collision, can be
obtained. Using the asymptotic properties of $C_{\rm cr}$ properly
scaled, Laskar obtains scaling laws for the spacing between adjacent
planets and their masses~{\mybf\citep{Laskar2000}}. We write 
Laskar's results as%
\begin{eqnarray}
  \label{eq:spacing}
  \delta_a/a^{1/2}&=& C_{\rm cr}^{1/3} \mu^{-1/6}
  \rho_a(a)^{-1/3} \delta n /k \\
  \label{eq:mass}
  m(a)&=& C_{\rm cr}^{1/3} \mu^{-1/6} a^{1/2} 
  \rho_a(a)^{2/3}/k .
\end{eqnarray}
Here, $\delta_a=a_{i+1}-a_i$ denotes the spacing between two nearby
planets ($\delta n=1$), $\rho_a(a)$ is the initial mass distribution
of planetesimals, $m(a)$ is the mass of the planet, and $k$ is a
dimensionless constant which depends on the ratio of the colliding
masses. Equation~\ref{eq:spacing} provides the spacing of the planets,
while Eq.~\ref{eq:mass} provides their masses; both are given in terms
of the initial mass distribution $\rho_a(a)$. We notice that $C_{\rm
  cr}$ in these equations is actually an unknown quantity.

{\mybf\citet{Laskar2000}} illustrated his findings numerically using a
constant linear mass-density distribution $\rho_a(a)=\zeta_a$ for the
planetesimals, a total mass of the disc $m_T=8\times10^ {-6}$ (in
solar masses) and for the AMD the value $C_{\rm
  final}=16\times10^{-8}$ as the statistically significant
quantity. His numerical results, in terms of the number of the planet
$n$, showed good agreement with respect to the analytical predictions,
given by Eqs.~\ref {eq:spacing} and~\ref{eq:mass}.

\subsection{Comments}

Laskar's model described above is elegant for its simplicity and predictions, 
regardless of the lack of any characteristic physical time-scale. 
Indeed, Eq.~\ref{eq:spacing} can be integrated explicitly
in many cases {\mybf\citep{Laskar2000}}. In particular, for
$\rho_a(a)=\zeta_a a^{z_a}$, we have that for $z_a\ne -3/2$ this
yields a power-law spacing distribution
$a_n^{z_a/3+1/2}-a_{0}^{z_a/3+1/2}\propto n$.  As an example, 
$z_a=0$ yields the square-root behaviour $a_n^{1/2}\sim n$.  Likewise,
for $z_a=-3/2$ it yields $a_n= a_0 \exp{[C_{\rm cr}^{1/3}\mu^{-1/6}
  n/(k\zeta_{a}^{1/3})]}$, which is basically the usual Titius--Bode
law~\citep{Nieto1972}.  Similar expressions can also be obtained for
the mass in terms of $n$ using Eq.~\ref{eq:mass}.

While the results are attractive, there is an important subtle point
in Eqs.~\ref{eq:spacing} and~\ref{eq:mass} as they stand. 
{\mybf As mentioned above, the total AMD is constant between consecutive 
collisions and decreases only when they occur. Then, below a critical
value $C_{\rm cr}$ it remains constant. Note that $C_{\rm cr}$ is 
not a constant of motion in the dynamical sense.
For the comparison with the numerical simulations, Laskar uses
the statistically significant quantity $C_{\rm final}$,
the final AMD. {\mybf While} $C_{\rm final}$ is not a constant of motion,
because different statistical realizations of the model yield
different values, the numerical results are quite satisfactory. }
Indeed, in order to achieve the fixed value $C_{\rm final}$, 
as a methodological procedure {\mybf\citet{Laskar2000}} excites 
the eccentricities of the planets {\mybf at the end of the simulation}, 
thus losing the constancy of the total angular momentum of 
the system; {\mybf note that it is here that Laskar's implementation fails to 
conserve the total angular momentum of the system.}
This observation is important since the conservation of
angular momentum is a cornerstone of the theories on planetary
formation and evolution~\citep{FernandezIp84}. Therefore, for the
integration of Eqs.~\ref{eq:spacing} and~\ref{eq:mass}, some previous
knowledge of the distribution of the $C_{\rm final}$ is required, or
at least its dependence with respect to other initial parameters,
e.g., $\rho_a(a)$. Note that knowledge of the distribution of $C_{\rm
  final}$ does provide means to consistently define $C_{\rm cr}$.

\begin{figure}
 \includegraphics[width=8.4cm]{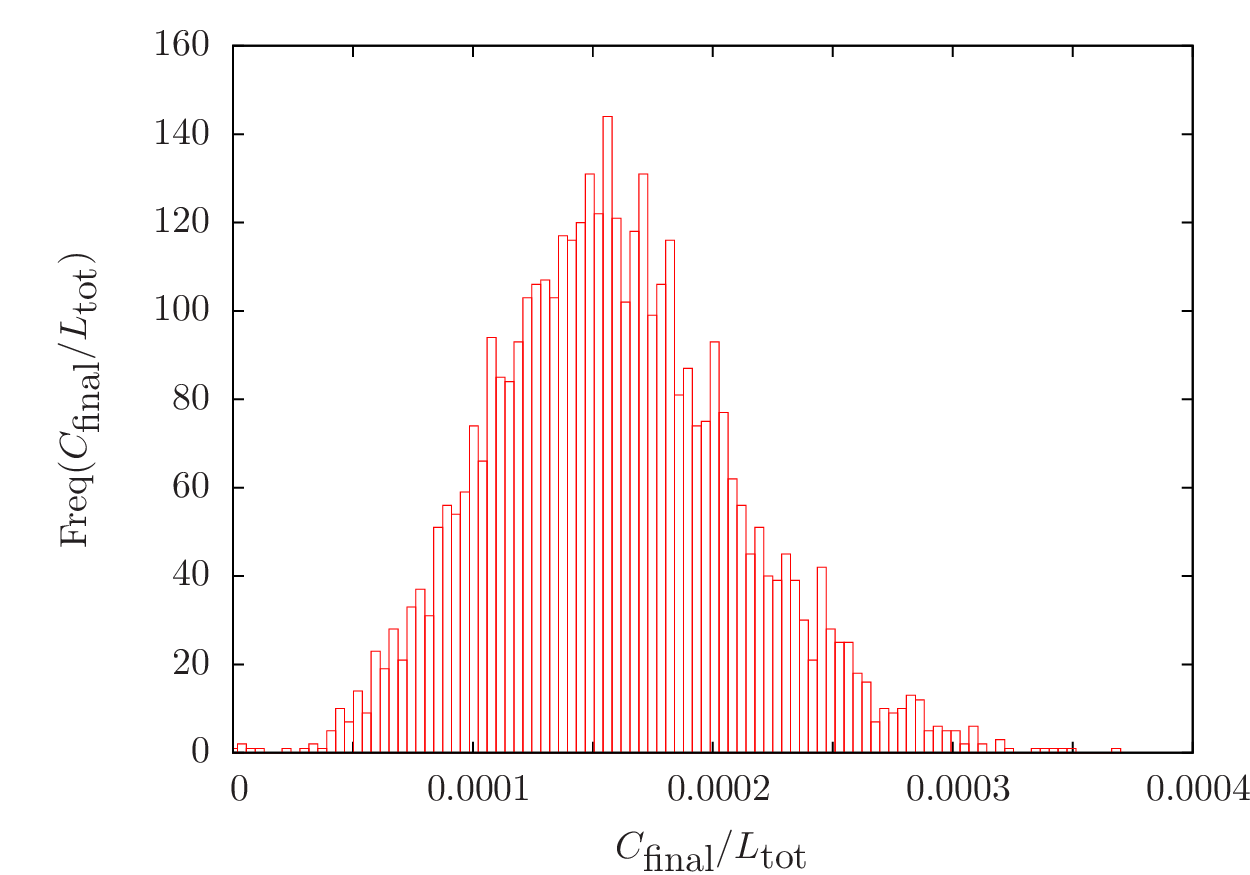}
 \caption{\label{fig1}%
   Frequency distribution of $C_{\rm final}/L_{\rm tot}$ for $4096$
   independent realizations of Laskar's model keeping fixed the total
   angular momentum of the system. In these calculations we fixed
   $L_{\rm tot} = 2.2212\times 10^{-2} M_\odot \, {\rm AU}^2 \, {\rm
     yr}^{-1}$, which is roughly the angular momentum of the planets
   of the Solar System. }
\end{figure}

To illustrate this point, in Fig.~\ref{fig1} we plot the frequency
distribution of the final AMD normalised to the total angular momentum
of the system, $C_{\rm final}/L_{\rm tot}$, for a large number of
realizations of Laskar's simplified model, keeping the total angular
momentum of the system constant. The figure clearly illustrates the
lack of constancy of the final AMD for different realizations. We
further observe that $C_{\rm cr}$ can be defined as the supremum of
$C_{\rm final}$; while this is a consistent definition, the numerical
calculation of such quantity may be involved due to the poor
statistics at the tail of the distribution.

{\mybf Aside from the lack of conservation} of $L_{\rm tot}$, in the
numerical implementation of Laskar's model, all relative orientations
of the elliptic trajectories of the colliding particles that intersect
lead to accretion with equal probability. In contrast, in the
derivation of Eqs.~\ref{eq:spacing} and~\ref{eq:mass}, Laskar imposes
a very specific collision among the particles. In either case, there
is no physical constraint with respect to the relative velocity of the
colliding particles, as stated by Safronov's
criterion~\citep{Safronov}. The latter establishes that accretion is
possible if the motion of the colliding particles satisfies $v_{\rm
  rel}\le v_{\rm esc}$, where $v_{\rm rel}$ is the relative velocity
of the colliding particles at the point of collision, and $v_{\rm
  esc}$ is the escape velocity of the local two-body problem. The
former depends --among other parameters defining the elliptic
orbits-- on the relative orientation $\theta$, while the latter can
be written as $v_{\rm esc}^2=2\mu/r_H$, where $r_H$ is the Hill radius
defined with respect to the dominating mass.  Moreover, the accretion
probability in Laskar's original model is independent of the orbital
and physical parameters of the particles, being in this sense
tantamount of an orderly type of
growth~\citep{Wetherill:1989Icarus77p792}. While this is certainly a
good starting point, it is intuitively important to include a
mass-dependence in the accretion probability, since it affects the
collision cross-section. A particular case of interest is the
so-called {\it runaway growth}~\citep{Greenberg1978}, during which the
more massive bodies grow faster than the lighter ones.

\section{Variations on Laskar's model}
\label{variations}

In this section we describe the variations and the numerical results
obtained by implementing some variations to Laskar's simplified
planetary model. These variations impose, firstly, the conservation of
the total angular momentum of the system, and secondly, address the
consequences of using mass-dependent accretion processes and velocity
selection mechanisms, not considered 
previously~{\mybf\citep{Laskar2000,LaskarPrep}},
without leaving the overall simplicity in the formulation and
implementation of the model. 

\subsection{General characteristics of the simulations}
For the numerical results presented in this section, we consider that
the mass of the host star is one solar mass ($M_0= 1 M_\odot$), which
we shall use as the unit of mass. We shall also fix the astronomical
unit (AU) as the unit of distance and the yr as the unit of time. In
these units we have $\mu={\cal G} M_0=4 \pi^2 \, {\rm AU}^3\,{\rm
  yr}^{-2}\, M_\odot^{-1}$. In addition, we shall fix the total mass
of the disc to $M_{\rm disc} = 1.3413\times 10^{-3} M_\odot$ and the
total angular momentum of the system to $L_{\rm tot} = 2.2212\times
10^{-2} M_\odot \, {\rm AU}^2 \, {\rm yr}^{-1}$. These values are
roughly the current values of mass and total angular momentum of the
planets in the Solar System, respectively. These parameters define the
physical quantities of our simulations.

In order to define the initial conditions, we must specify the form of
the initial linear mass-density distribution $\rho_a(a)$ (or
equivalently the initial surface density $\rho_a(a)/a$) and the
initial eccentricity distribution $\rho_e(e)$. These distributions and
the extension of the disc are constrained by the total mass and the
total angular momentum of the system. In the continuum limit, the
total angular momentum of the system is written as
\begin{equation}
  \label{eq:lztot}
  L_{\rm tot} = \int_{a_{\rm min}}^{a_{\rm max}} \int_0^{e_{\rm max}} \rho_a(a) 
    \rho_e(e) \sqrt{\mu a (1-e^2)}\, da\,de.
\end{equation}
Assuming that $a_{\rm min}$ and $e_{\rm max}$ are given quantities,
Eq.~\ref{eq:lztot} can be used to define the maximal extension of the 
initial disc, $a_{\rm max}$ . Notice that Eq.~\ref{eq:lztot} can be
explicitly integrated for a homogeneous eccentricity density
$\rho_e(e)=\zeta_e$ and a power-law linear mass-density distribution,
$\rho_a(a)=\zeta_a a^{z_a}$~\citep{MenaPhD}.

For a fair comparison with Laskar's results {\mybf\citep{Laskar2000}}, our 
simulations 
are initiated with a large number (typically 10000 bodies) of equal mass
planetesimals, which is the most constraining case in terms of equivalence 
of the planetesimal location~\citep{Namouni1996}; similar results are obtained
for a larger initial number of planetesimals. We also
considered the planetesimals on coplanar orbits, defining their initial
semi-major axes and eccentricities from a homogenous linear
mass-density distribution $\rho_a(a)=\zeta_a$ and a homogeneous
eccentricity distribution $\rho_e(e)=\zeta_e$, respectively. Using
$a_{\rm min}=0.1$~AU and $e_{\rm max}=0.3$ in Eq.~\ref{eq:lztot} yields
for the maximal extension of the disc $a_{\rm max}\cong
15.93$~AU. This value of $a_{\rm max}$ is smaller (by a factor 2) than
the $30-35$~AU estimate for the outer edge of the original
planetesimal disc of the Solar System, proposed by
\citet{Levinson2003}. However, this value is fully consistent with the
range for the initial conditions used for the giant planets in the
simulations of \citet{Tsiganis2005}. The contributions of the massive
disc ($30-50$~M$_{\oplus}$) that these authors include beyond the
orbits of the planets up to $30-35$~AU are excluded from the values
$M_{\rm disc}$ and $L_{\rm tot}$. Notice that the outer disc has also a constant
linear mass-density~\citep{Tsiganis2005}. Finally, we emphasize that the use of 
Eq.~\ref{eq:lztot} to define the maximal extension of the initial disc,
$a_{\rm max}$, is yet another difference with respect to Laskar's
implementation.

It is also worth describing the implementation of the Brownian motion
that accounts for the chaotic evolution of the averaged
equations. Instead of implementing it in the space of eccentricities,
we accomplished this by the exchange of angular momentum between a
pair of arbitrary planetesimals. Denoting the initial angular momenta
of the chosen planetesimals by $l_1$ and $l_2$ and the corresponding
final ones as $l_1^\prime$ and $l_2^\prime$, we write
$l_1^\prime=l_1+\delta l$ and $l_2^\prime=l_2-\delta l$. This ensures
the conservation of the total angular momentum. The angular momentum
exchanged is written as $\delta l=\beta \xi$. Here, $\beta\le 1$ is a
positive dimensionless parameter related to the diffusion constant; in
our simulations we considered the value $\beta=10^{-3}$ which is small
enough to model the slow secular chaotic
diffusion~\citep{Laskar1989,Laskar1990}.  The random walk character is
provided by the stochastic variable $\xi$, which is a uniformly
distributed random number in the interval
$[-c_2^\prime,c_1^\prime]$. Here, $c_1^\prime$ and $c_2^\prime$ are
the largest positive numbers (with units of angular momentum) that
simultaneously satisfy the inequalities $c_1^\prime\le C_1$,
$c_1^\prime\le l_2$, $c_2^\prime\le C_2$, $c_2^\prime\le l_1$, where
$C_1$ and $C_2$ are the AMD's of the planetesimals. This definition of
$\delta l$ ensures that both $l_1^\prime$ and $l_2^\prime$ lie in the
correct intervals and the new eccentricities are well-defined. Note
that this definition permits collisions of very eccentric ($e_i\sim
1$) planetesimals with the host star~\citep{Weidenschilling1975}.
This is consistently taken into account in our simulations whenever
the perihelion of a particle is smaller than the radius of the Sun
($4.65\times 10^{-3}$~AU).

\subsection{Laskar's model including angular momentum
  conservation: orderly growth}
\label{variation0}

We shall discuss firstly the case where we impose consistently the
conservation of the total angular momentum within the simplified model
of Laskar. This is aimed to clarify the differences that arise with
the variations we implement with respect to Laskar's results. As
mentioned above, the simulations end once the system satisfies the AMD
stability criterion. We carry out accretion and orbit evolution by
selecting at random a planetesimal and one of its neighbours for
accretion, and then proceed with the exchange of angular momentum
between (many) pairs of planetesimals, also chosen at random. The mass
distribution evolves smoothly, which makes this case similar to an
{\it orderly growth}~\citep[cf.][]{Armitage:2007arXiv}.

\begin{figure}
  \includegraphics[width=8.4cm]{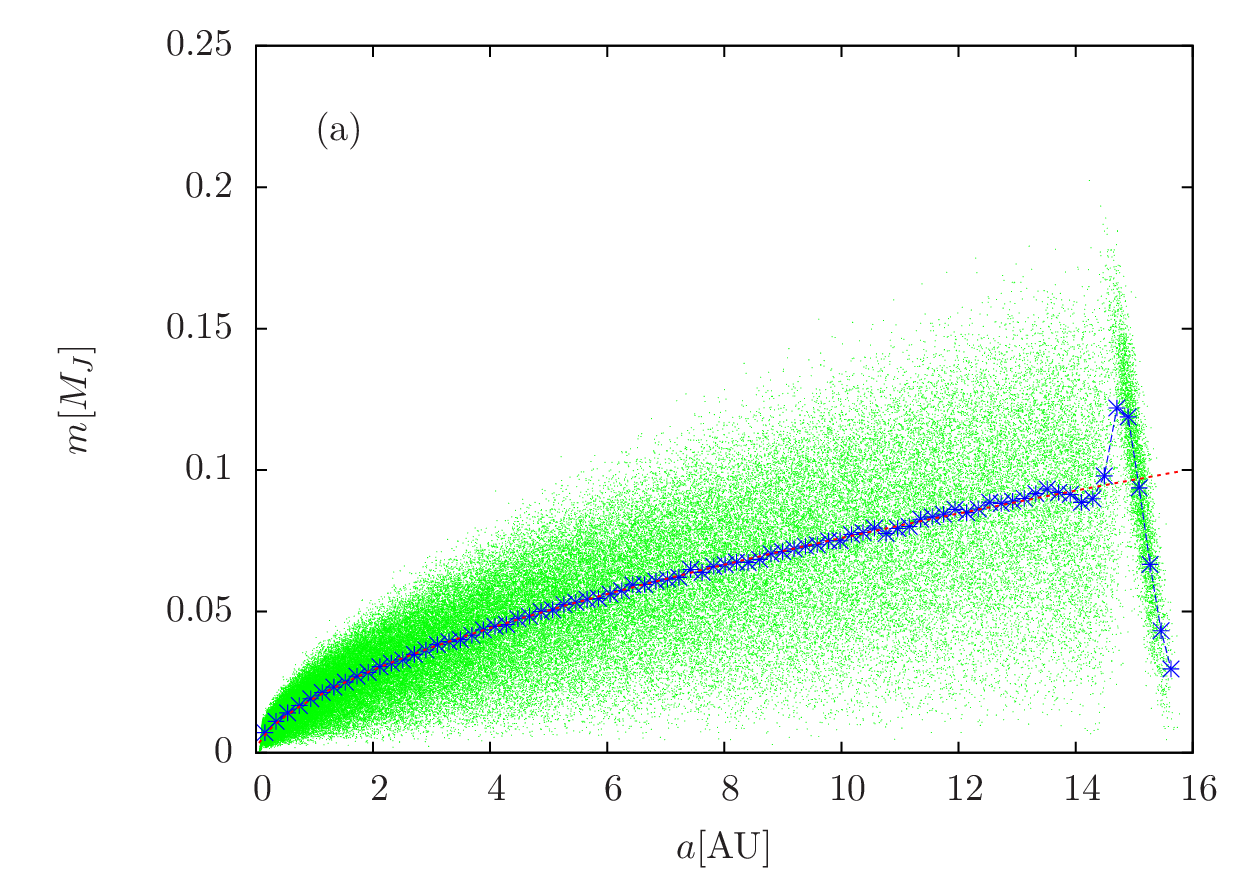}\\
  \includegraphics[width=8.4cm]{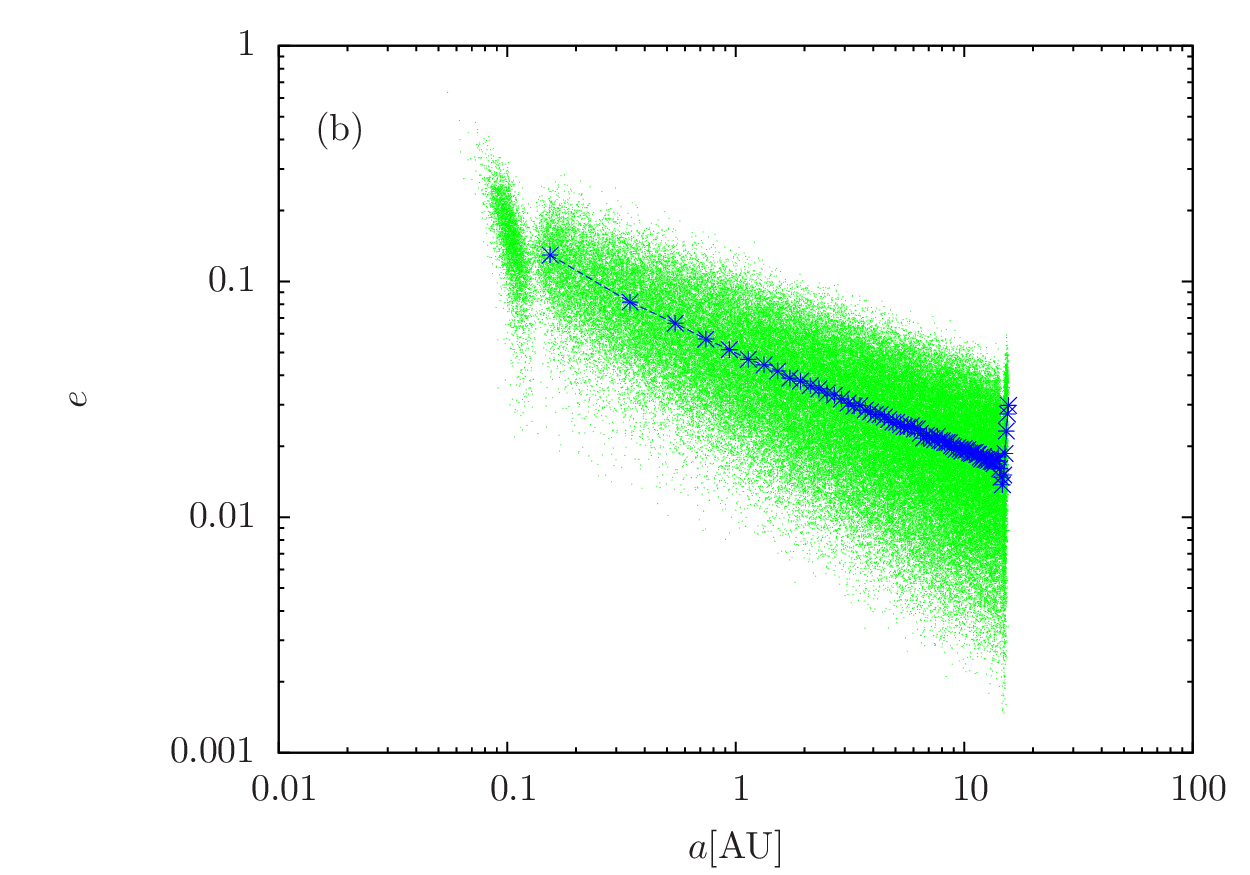}
  \caption{\label{fig2}%
    (a)~Final planetary mass and (b)~eccentricity in terms of the
    corresponding semi-major axis for $4096$ realizations of Laskar's
    simplified model, imposing the conservation of the total angular
    momentum of the system. Each dot represents a planet formed in one
    of the simulations. The star symbols represent local averages
    taken over a semi-major axis interval of $0.2$~AU. In the top
    panel, the curve represents the power-law fit
    (Eq.~\ref{eq:massfit}) of the averaged data in the interval
    $a\in[0,14.5]$. }
\end{figure}

\begin{figure}
  \includegraphics[angle=0,width=8.4cm]{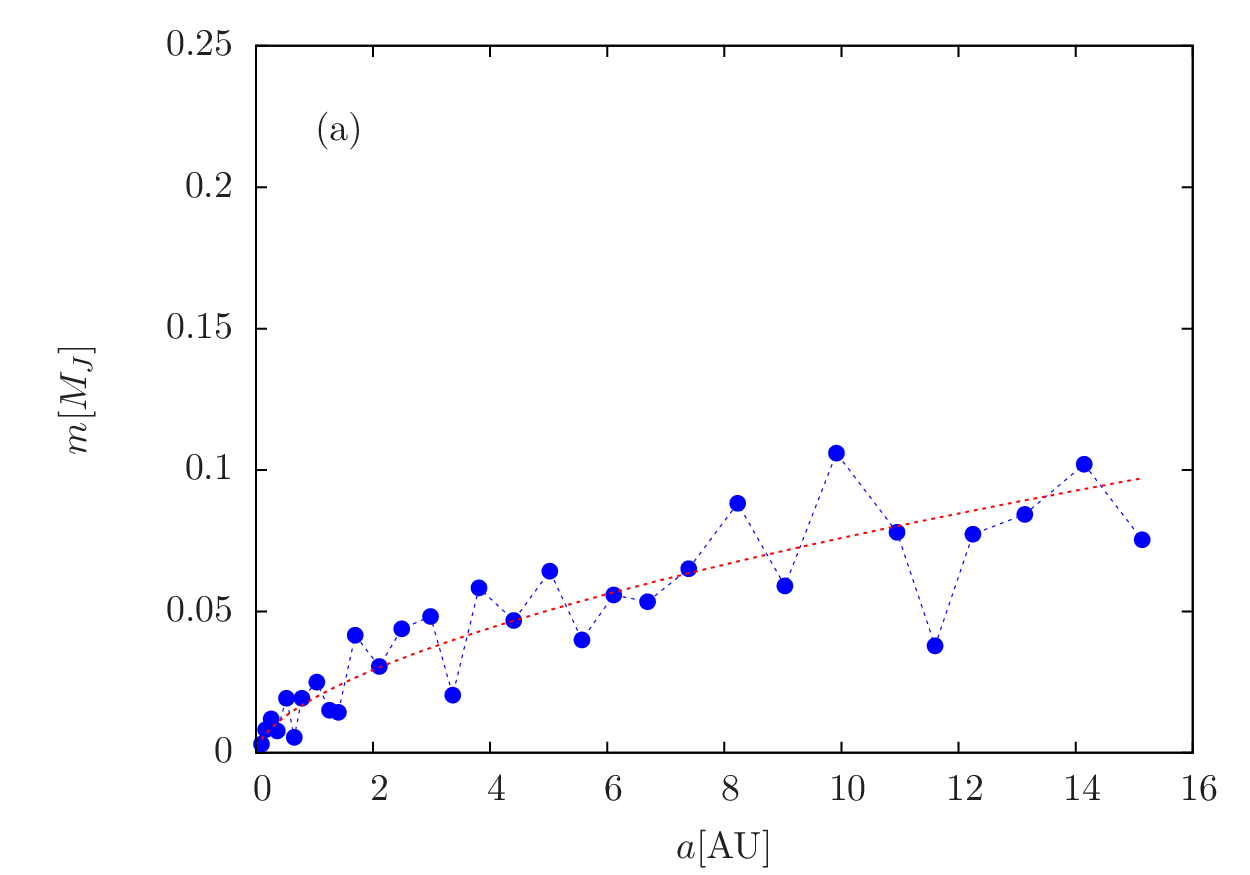}
  \includegraphics[angle=0,width=8.4cm]{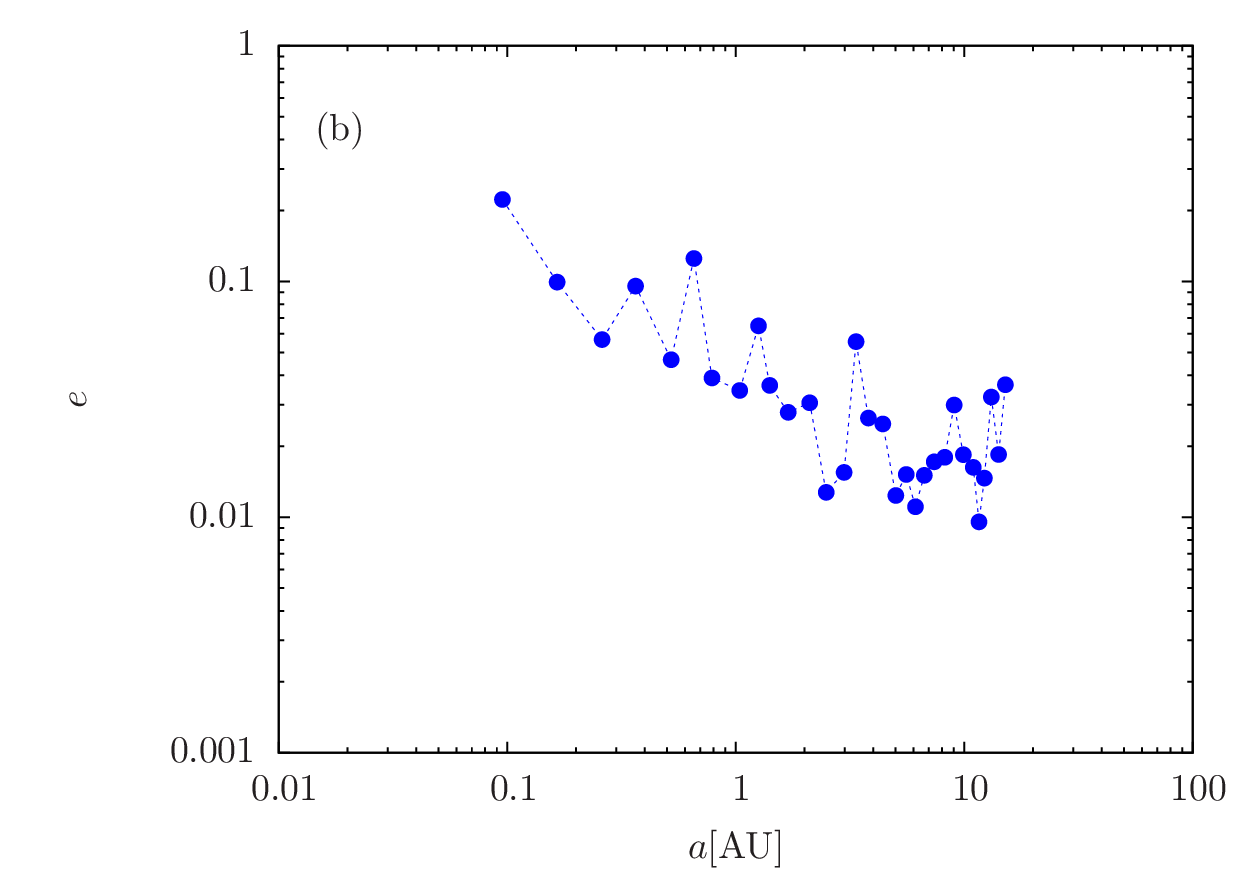}
  \caption{\label{fig3}%
    (a)~Final planetary mass and (b)~eccentricity in terms of the
    semi-major axis for a typical (randomly chosen) planetary system,
    obtained by implementing Laskar's simplified model with
    conservation of the total angular momentum (orderly-growth model).
    The continuous curve in the top panel is the the power-law fit
    (Eq.~\ref{eq:massfit}) obtained from the averaged data of the
    ensemble of Fig.~\ref{fig2}(a).}
\end{figure}

In Fig.~\ref{fig2} we present the final planetary configurations by
combining the results of all $4096$ realizations of the model,
plotting the mass and eccentricity of the formed planets in terms of
their semi-major axis. Figure~\ref{fig3} illustrates the resulting
final configuration of one planetary system taken at random from the
set of realizations. In both figures (and throughout this section),
the final planetary masses are given in units of the mass of Jupiter
$M_J$ ($1 M_\odot=1047.56 M_J$; $1 M_J=317.83 M_{\oplus}$).

The results presented in Fig.~\ref{fig2}(a) show an increasing trend
of the planetary masses up to $a\approx 14.5$~AU, where a rapid
decrease of $m$ in terms of $a$ takes place. The latter is due to the
finite size of the planetesimal disc and the conservation of the
angular momentum, and corresponds to the location of the outermost
planet of the planetary systems formed. We observe that the largest
planetary mass attained is $\sim 0.2 M_J$, which is a comparatively
small value with respect to the mass of Jupiter. This is a consequence 
of the large number of planets
formed in each simulation, producing an average of $30.9\pm1.7$
planets (the error is the standard deviation).

As illustrated in Fig.~\ref{fig3}, individual planetary systems may
not display the purely initial increasing behaviour of $m(a)$ observed
in the data of the ensemble. The results of the ensemble are
statistical and it is thus meaningful to define the {\it average behaviour} 
of the ensemble. {\mybf\citet{Laskar2000}} computes the average
of the final planetary mass and semi-major axis for fixed planet
number $n$. Since $n$ is not an observable quantity, we have opted to
average the results over a small interval of the semi-major axis,
which was fixed to $0.2$~AU. The resulting (binning) average is
represented by the star symbols in Fig.~\ref{fig2}. For comparison
with Eq.~\ref{eq:mass}, we fit the averaged data with the power-law
function
\begin{equation}
  \label{eq:massfit}
  m(a) = A a^\nu.
\end{equation}
According to Eq.~\ref{eq:mass} we should expect $\nu=1/2$. The
continuous curve in Fig.~\ref{fig2}(a) illustrates the resulting
best-fitting, which in this case was calculated up to $a=14.5$~AU
where the decreasing branch of the last planet sets in; the best-fitting
yields $\nu\approx 0.59$. This value of $\nu$ is close but larger than
Laskar's predicted value $1/2$. We attribute this difference to the
(implicit) dependence of the distribution of $C_{\rm cr}$ on the
specific parameters of the model; this dependence is not considered in
the integration of Eqs.~\ref{eq:spacing} and~\ref{eq:mass}. In
Fig.~\ref{fig3}(a) we also plot the best-fitting curve, in order to
compare the results of a specific random planetary system with the
results averaged over the ensemble.

Figure~\ref{fig2}(b) displays the results of the final planetary
eccentricity for the ensemble on a logarithmic scale, and Fig.~\ref{fig3}(b) 
the corresponding results for a typical member of the ensemble. We observe a
discontinuous behaviour around $a\sim 0.1$~AU. This feature is similar
to the discontinuity related to the location of the {\mybf outermost} planet
discussed in the mass diagram; in this case, it indicates the
localisation of the innermost planets. The figure shows that the final
eccentricities are smaller for larger semi-major
axis~\citep{Jones2004}. Planets with smaller masses display larger
eccentricities in comparison with planets with larger masses. This is
a consequence of the accretion processes, which tend to circularise
the planetary orbits, consequently causing the more massive planets to
exhibit smaller eccentricities than those less massive.

\subsection{Mass-selection mechanism: runaway growth}
\label{variation1}

As mentioned earlier, in Laskar's original model the accretion
probability is independent of the mass of the planetesimals. Yet, mass
increases the collisional cross-section and thus promotes accretion,
at least once the gravitational attraction among planetesimals starts
to dominate their dynamics. We consider now the inclusion of such
effects in the model. We address the case where the massive bodies
grow quite fast due to their mass, and thus increase their difference
in mass with respect to the lighter bodies. This happens until they
exhaust the available mass in their surroundings. In what follows, we
shall refer to this process as {\it mass-selection} or {\it runaway
  growth}~\citep{Greenberg1978,Wetherill:1989Icarus77p792}. Notice
that, while we are including now the mass to select the particle that
accretes, the relative velocity of the colliding particles is not yet
taken into account, which is another important property that enhances
gravitational focussing~\citep{Armitage:2007arXiv}.

We implement the mass-selection growth as follows. At each iteration
of the model we define a threshold mass $m_{\rm cut}$, which is a
uniformly-distributed random number between zero and the mass of the
heaviest planetesimal. We then select at random a planetesimal whose
mass is larger than this threshold mass, and implement the accretion
process with one of its randomly-selected neighbours; the relative
orientation of the elliptic orbits of the colliding particles is
considered as before. With this simple implementation, once a particle has 
a mass slightly larger 
than the rest, i.e., from the first iteration, the probability that this particle 
is selected for collision in the next iteration is dramatically increased in 
comparison to other particles. This promotes accretion of the more massive 
particles, and hence mass growth in a runaway-like form. Additionally, it 
will systematically open a gap around the location of the accreted particle, 
since the distance to the nearest neighbours increases. Eventually, the 
nearest particles will be far enough to avoid collisions with this particle, i.e., 
the geometric condition~\ref{eq:colision} is not satisfied. In this case the 
planetesimal becomes locally isolated, until another particle is close 
enough again or enough angular momentum has been exchanged
to allow for a new collision. If the particles that satisfy the condition
$m_i>m_{\rm cut}$ can not collide with their neighbours (the isolation mass 
is locally
reached), we then reset the value of $m_{\rm cut}$ to a smaller value,
and proceed as described above. Clearly, this implementation promotes
accretion of the more massive planetesimals. The chaotic diffusion
following the accretion processes is implemented as before. Again, the
simulations are iterated until the condition of AMD stability is
reached.

\begin{figure}
  \includegraphics[width=8.4cm]{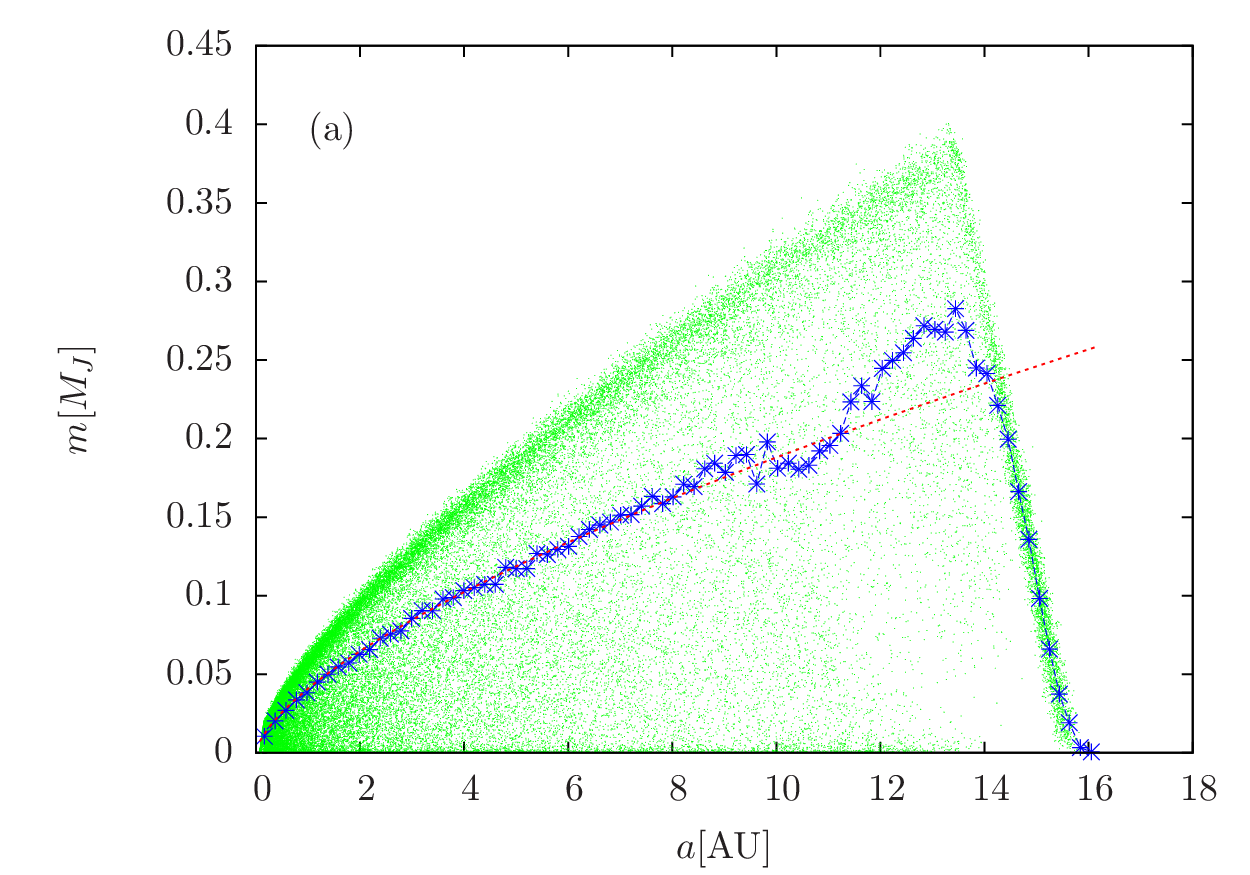}\\
  \includegraphics[width=8.4cm]{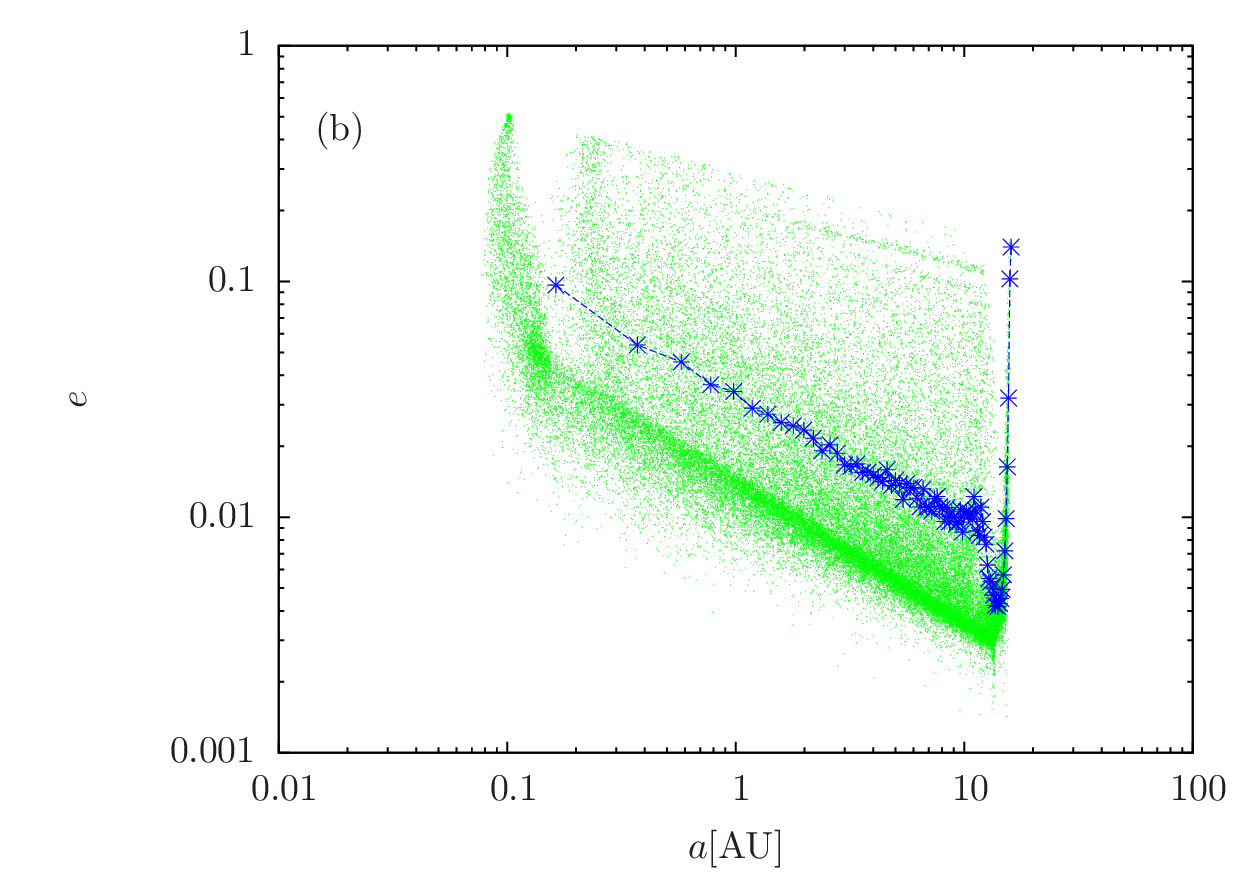}
  \caption{\label{fig4}%
    Same as Fig.~\ref{fig2} combining $4096$ realizations of the model,
    when the implementation includes runaway (mass-selection) growth. In
    the top panel, the fit with the power-law Eq.~\ref{eq:massfit} 
    of the averaged data, represented by the continuous curve, was 
    carried in the interval $a\in[0,11.4]$.}
\end{figure}

\begin{figure}
  \includegraphics[angle=0,width=8.4cm]{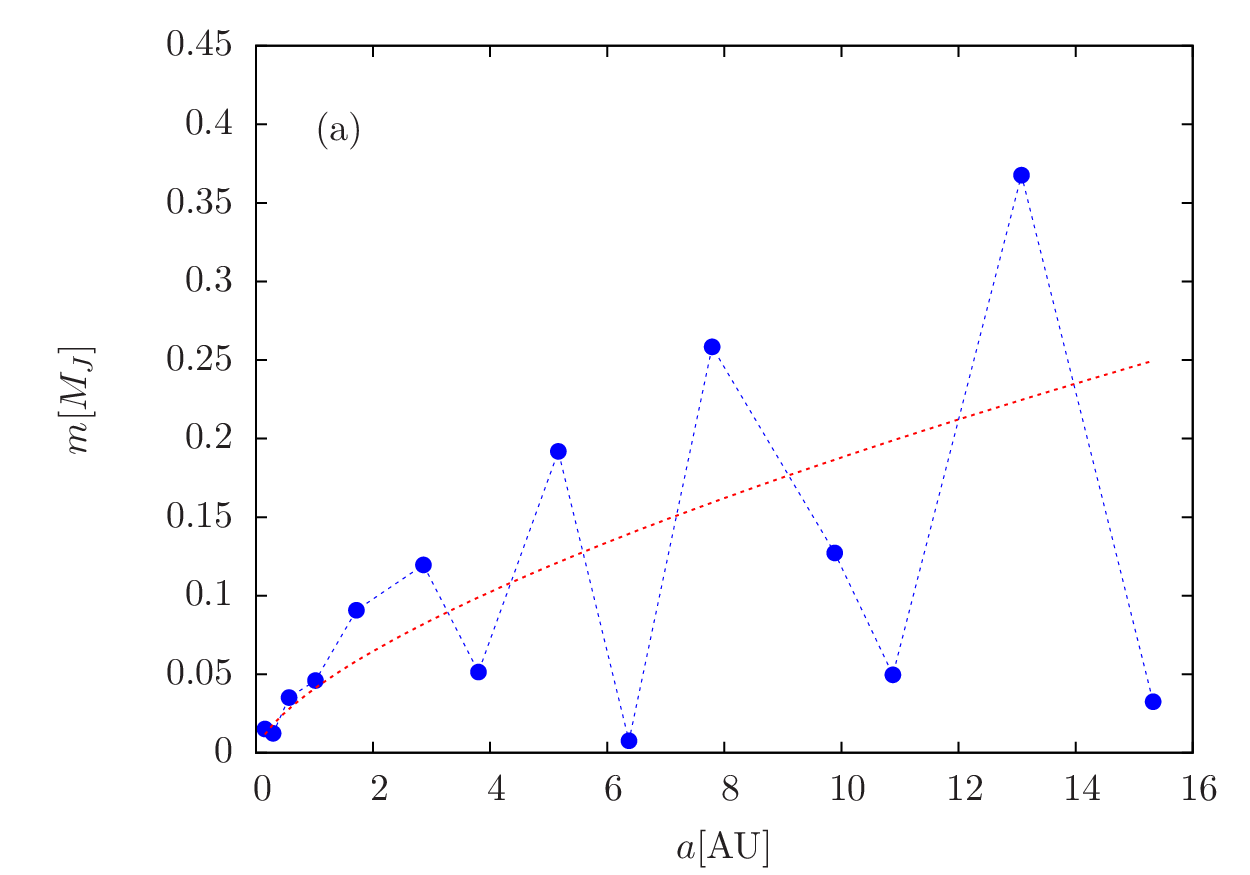}
  \includegraphics[angle=0,width=8.4cm]{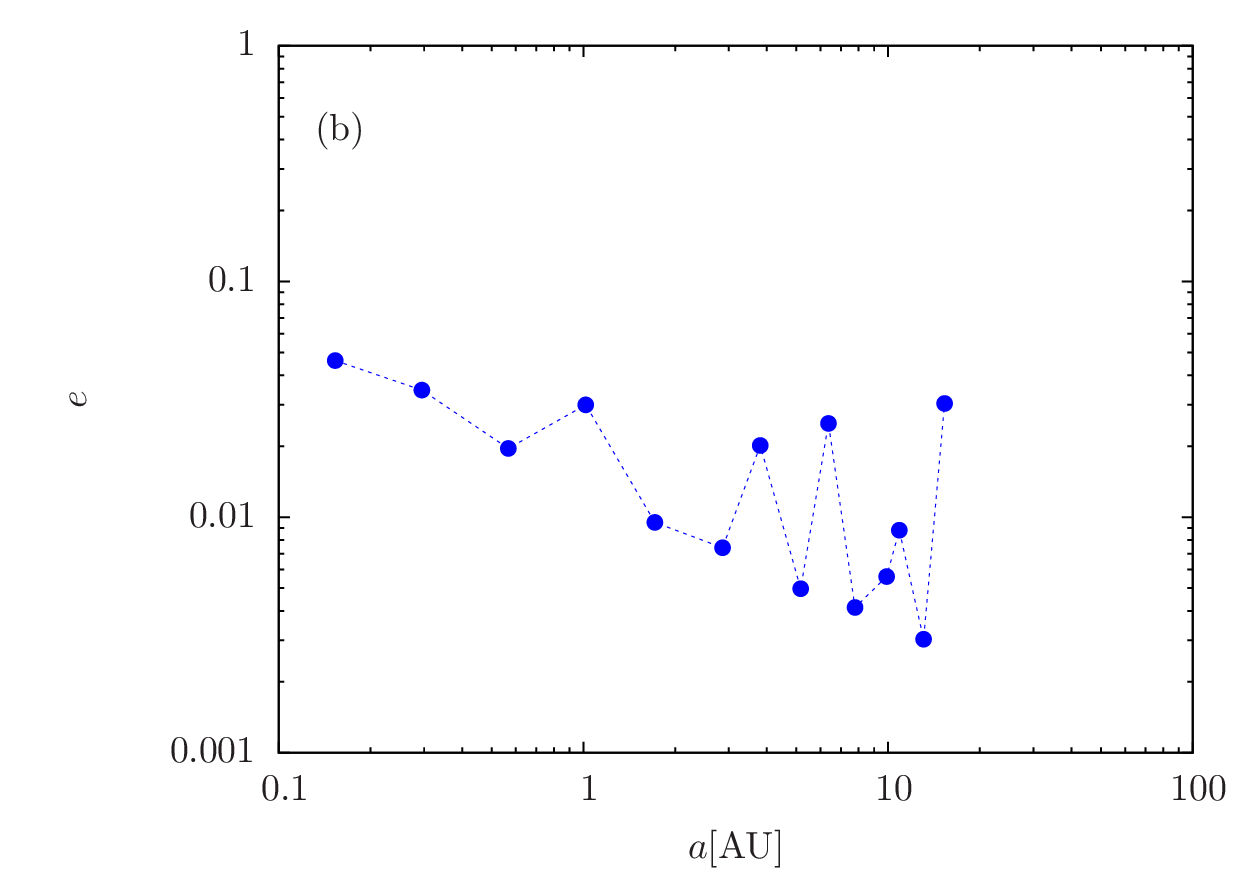}
  \caption{\label{fig5}%
    Same as Fig.~\ref{fig4} for one typical planetary system obtained 
    in the runaway (mass-selection) growth model. The continuous curve
    in the top panel corresponds to the power-law fit of the averaged 
    data in Fig.~\ref{fig4}(a).}
\end{figure}

In Fig.~\ref{fig4} we present the results of the runaway
(mass-selection) growth model, illustrating the combined results of
the masses and eccentricities of the final planetary configurations in
terms of their semi-major axes, for $4096$ independent realizations of
the model.  Figure~\ref{fig5} displays similar results for a single
planetary system chosen at random. As illustrated by the results of
the ensemble, the mass increases with the semi-major axis up to
$a\approx 13.4$~AU, where a sudden decrease of the mass takes
place. This is similar to the case of orderly growth, where the
decreasing branch marks the location of the {\mybf outermost} planet formed. As
done before, we fit the local averaged mass (up to $11.4$~AU) with the
power-law function (Eq.~\ref{eq:massfit}).  This yields the exponent
$\nu\approx 0.66$, which is larger than the one obtained for the
averaged data of Fig.~\ref{fig2}(a). We have chosen to fit up to
$a=11.4$~AU because, beyond this value, the averaged mass displays a
strong oscillation. Notice that the overall mass-scale is enhanced by
a factor close to $2$ with respect to the orderly-growth case, but the
largest accreted particles are at most $0.4 M_J$. The average final
number of formed planets is reduced to $14.6\pm 1.15$.

It is interesting to compare the density of points in the
mass--semi-major-axis diagram. For a given small interval of $a$, in
Fig.~\ref{fig2}(a) the density of the ensemble of planets is quite
uniform with respect to the mass they reach (except towards the upper
edge). This statement holds essentially for all $a$ before the
last-planet branch appears. On the other hand, the density of points
in Fig.~\ref{fig4}(a) is manifestly non-uniform, at least for
semi-major axis larger than $\sim 2$~AU. Indeed, considering a small
interval of semi-major axis, there is a marked clustering of points
towards the upper and lower edges of the mass. According to these
observations, this runaway-growth mechanism yields larger final
planetary masses, and thus a smaller number of planets, but also a
systematic occurrence of small-mass planets for essentially all values
of $a$. Since these planetary systems are AMD stable, these small-mass
planets are far from planetary collisions. In this case, possible
mean-motion resonances not included in the model could increase their
eccentricities and eventually yield new collisions (see
e.g.~\citet{LaskarGastineau2009}).

The eccentricities of the planetary ensemble obtained in this case,
illustrated in Fig.~\ref{fig4}(b), differ also from those resulting
from orderly growth, cf. Fig.~\ref{fig2}(b). Figure~\ref{fig4}(b)
shows the lack of uniformity in the density of points; yet, we note
there is a density concentration towards comparatively smaller values
of the eccentricity. That is, comparing the average value of the
eccentricity for a given interval around $a$ for orderly and runaway
growth, in the latter the average eccentricity is smaller, except for
large values of $a$. Therefore, the mass-selection mechanism induces
more circularised orbits of the planets, and in this sense, the final
configurations are more stable.  We also note that in the present
case, individual planets typically with small $a$, may reach
comparatively larger values of the eccentricity.

\begin{figure}
  \includegraphics[width=8.4cm]{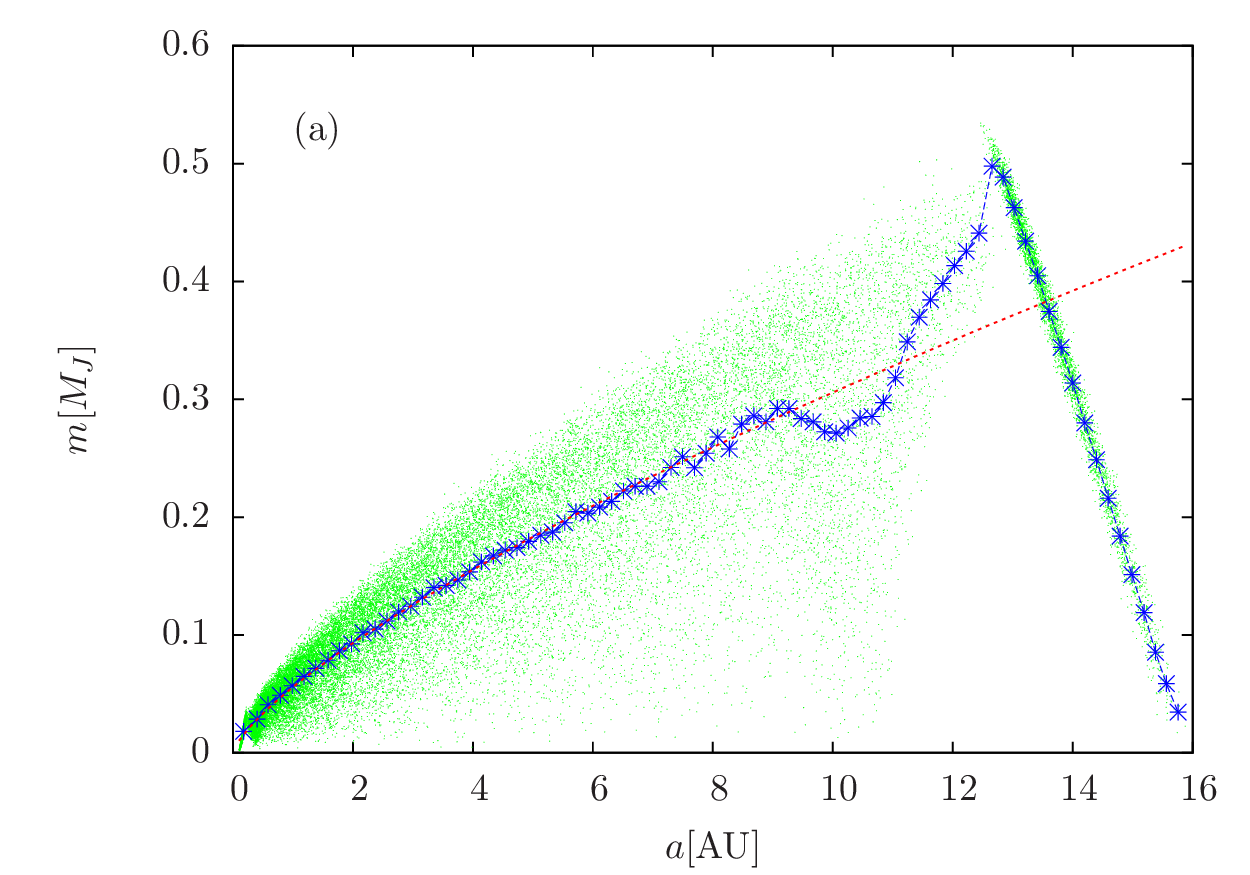}\\
  \includegraphics[width=8.4cm]{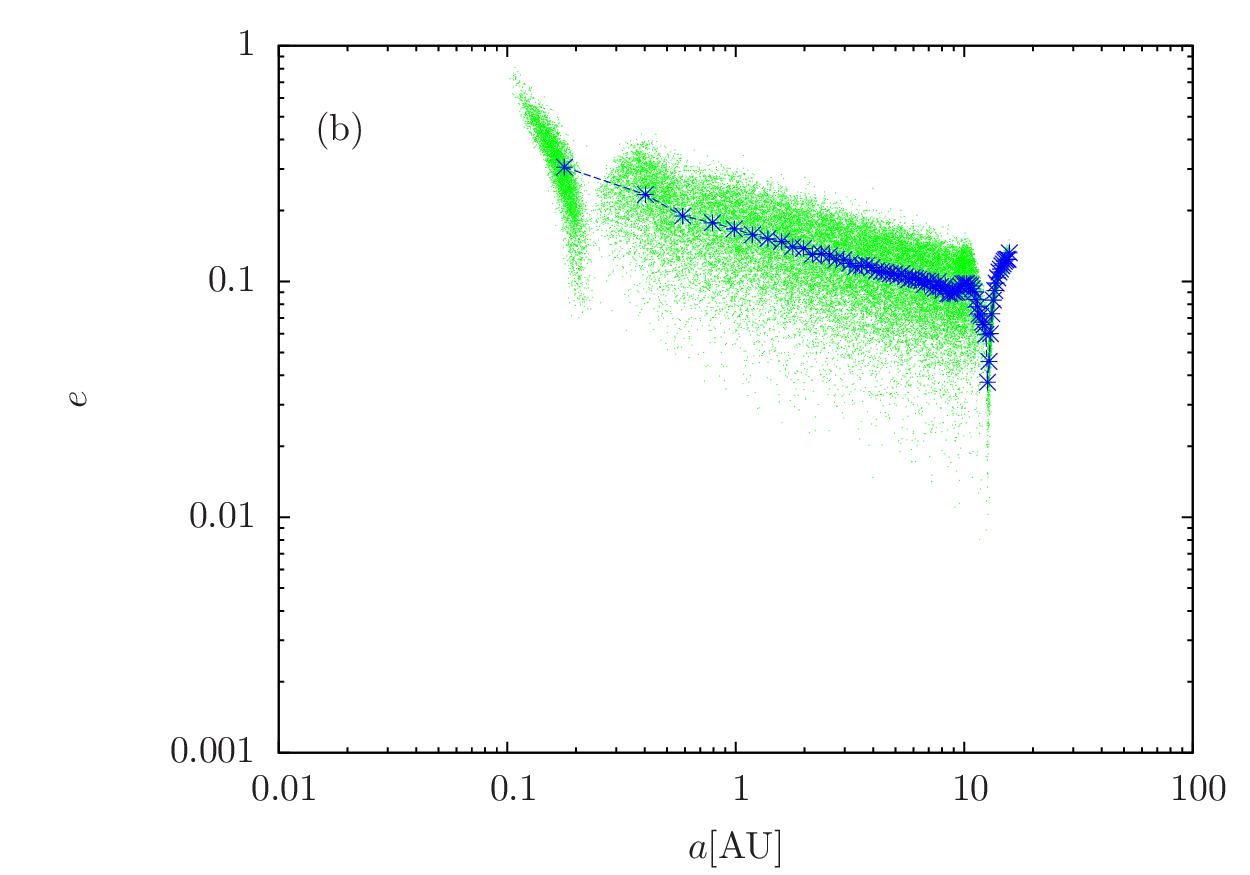}%
  \caption{\label{fig6}%
    Same as Fig.~\ref{fig2} for the combined results of $4096$
    simulations of the velocity-selection accretion model. In the top
    panel the continuos curve represents the fit of the power-law
    carried out in the interval $a\in[0,9.34]$.}
\end{figure}

\begin{figure}
  \includegraphics[width=8.4cm]{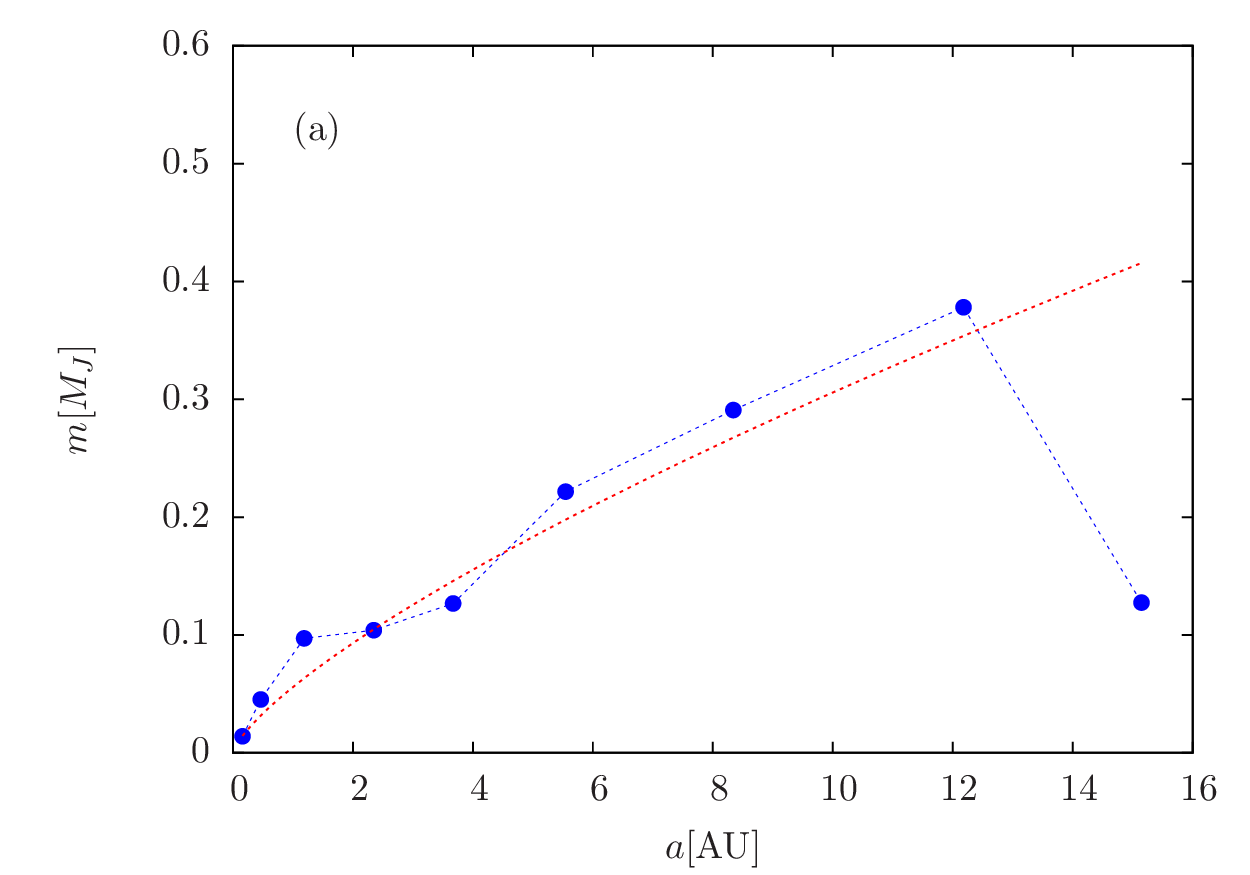}\\
  \includegraphics[width=8.4cm]{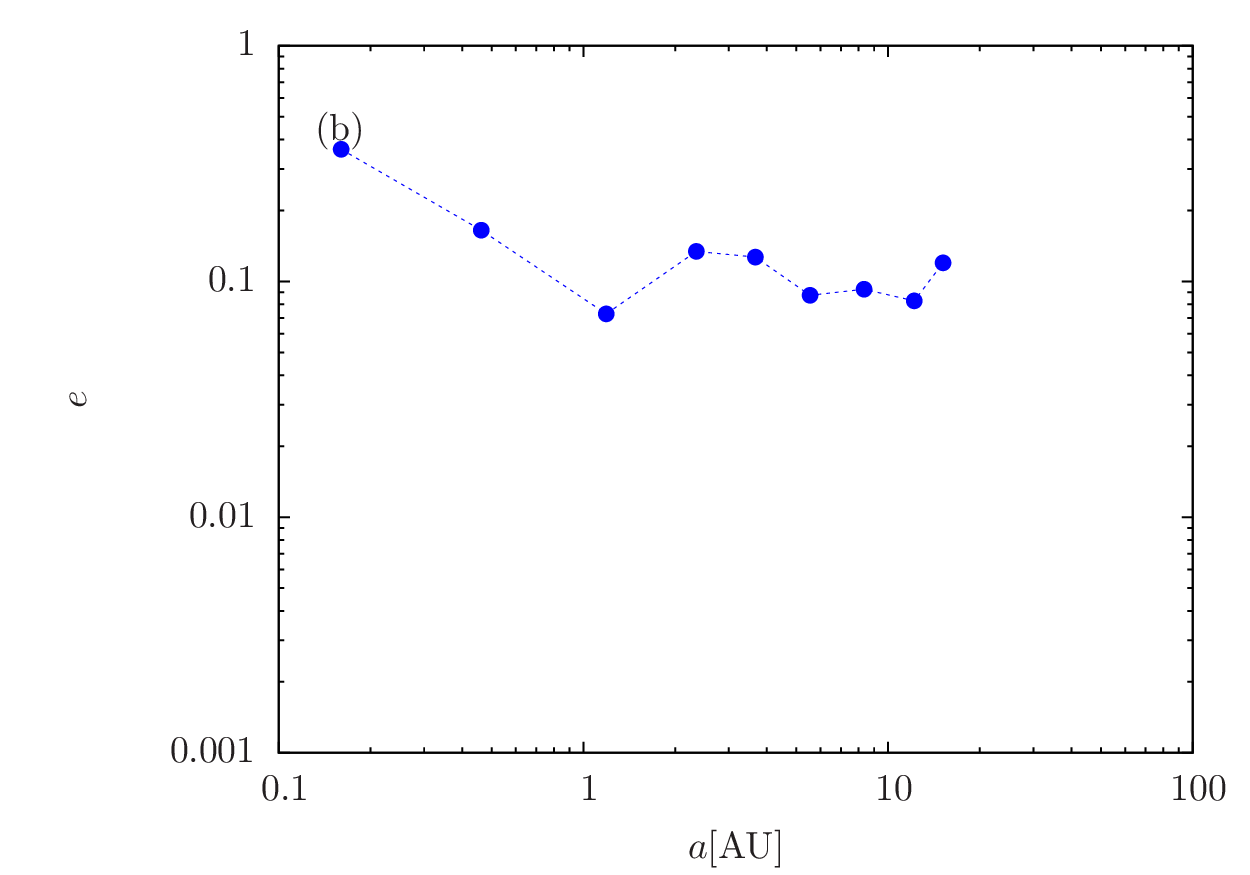}
  \caption{\label{fig7}%
    Same as Fig.~\ref{fig6} for a random planetary system obtained in
    the velocity-selection model. The continuous curve in the top
    panel is the power-law fit of the averaged data displayed in
    Fig.~\ref{fig6}(a).}
\end{figure}

\subsection{Velocity-selection mechanism}
\label{variation2}

Another important factor that influences the accretion probability is
the ratio between the relative velocity of the colliding bodies and
their escape velocity. In particular, this quantity enters in the
collisional cross-section through the gravitational focusing
factor~\citep{Wetherill:1989Icarus77p792} 
\begin{equation}
\label{eq:GravFocus}
F_g = 1 + v_{\rm esc}^2/v_{\rm rel}^2. 
\end{equation} 
Here, $v_{\rm rel}$ is the relative velocity of the colliding
particles (at infinity), and $v_{\rm esc}$ is the escape velocity in
the local two-body problem. For $v_{\rm rel}/v_{\rm esc} \ll 1$ the
gravitational factor enhances the collision cross-section. Moreover,
in this case the colliding bodies are expected to be bound together
and thus promote planetary growth~\citep[see][]{Armitage:2007arXiv}.

One naive way of implementing these effects into Laskar's simplified
model with the conservation of angular momentum, is by selecting the
relative orientation of the ellipses of the colliding bodies, such
that the condition $v_{\rm rel}/v_{\rm esc} \ll 1$ is satisfied. That
is, we allow accretion of particles only when Safronov's condition is
fulfilled. Carrying on these simulations resulted in systems where the
AMD stability condition was not fulfilled, in general. Therefore, the
elliptic trajectories of nearby bodies may intersect and thus collide,
but for those collisions no value of $\theta$, the relative
orientation of the ellipses, exists that $v_{\rm rel}/v_{\rm esc} \ll
1$ is fulfilled.

We therefore relaxed Safronov's condition for accretion, and fixed
$\theta$ as the relative orientation of the ellipses where $v_{\rm rel}$ 
achieves its minimum value.  By doing this, we promote a
moderate enhancement of the collisional cross-sections by the
gravitational focussing factor, as well as quasi-tangential
collisions. The latter is physically plausible for
accretion~\citep{Ohtsuki1993}. We note that with this implementation,
the change in the absolute value of the energy after accretion is
minimal, cf. Eq.~\ref{eq:energy}, thus minimising the net migration of
the accreted particle. We shall refer to this mechanism as the {\it
  velocity-selection mechanism}.

\begin{figure}
  \includegraphics[width=8.4cm]{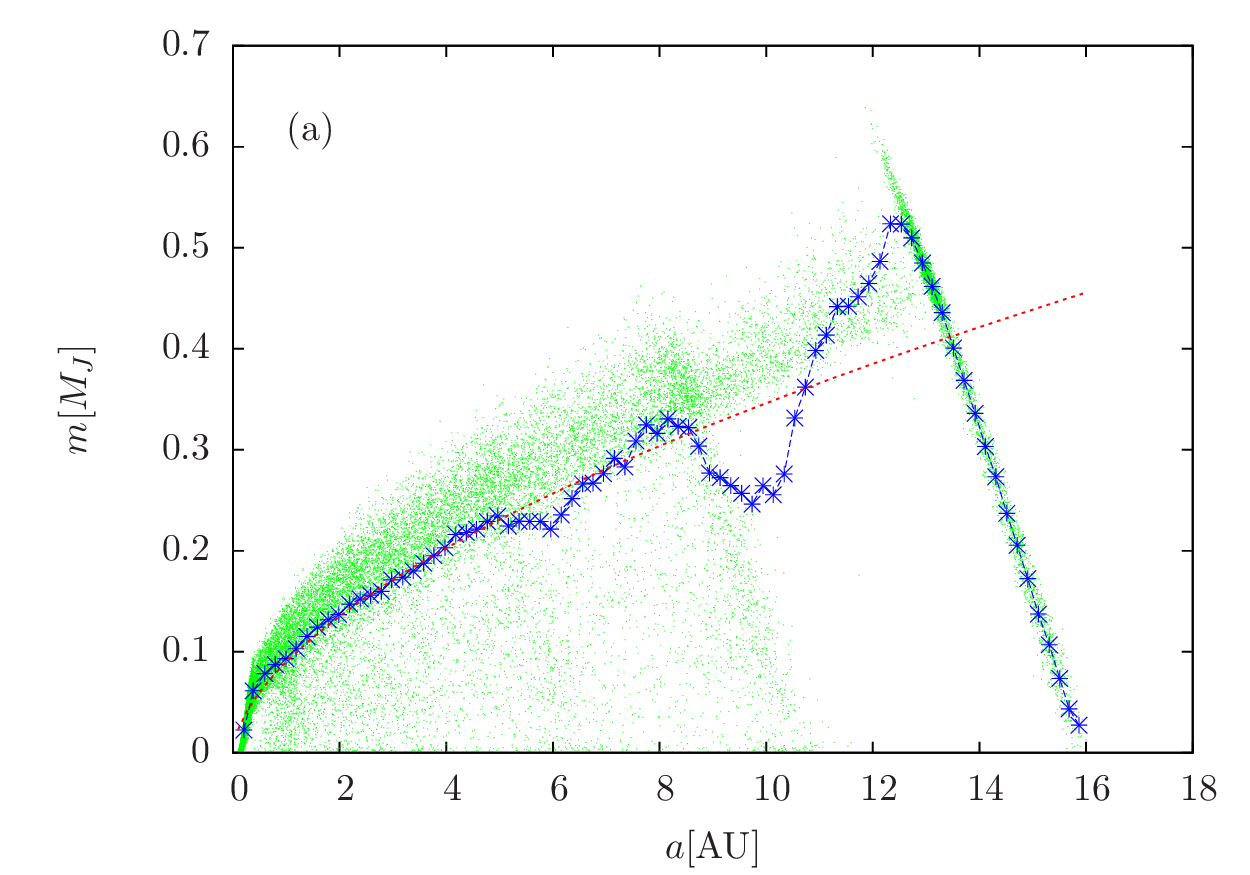}\\
  \includegraphics[width=8.4cm]{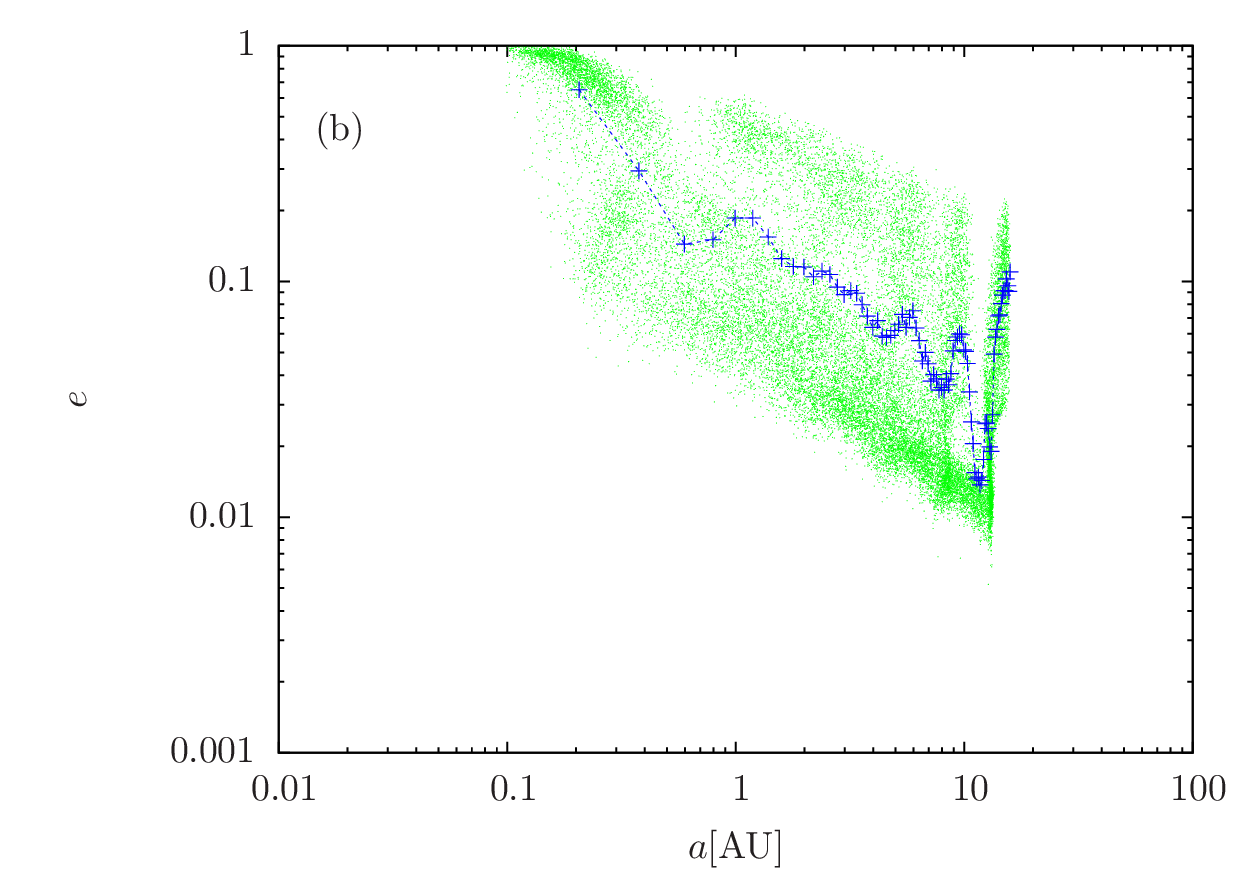}%
  \caption{\label{fig8}%
    Same as Fig.~\ref{fig2} for the combined results of $4096$
    simulations of the oligarchic growth model. In the top panel the
    continuos curve represents the fit of the power-law carried out in
    the interval $a\in[0,8.6]$.}
\end{figure}

\begin{figure}
  \includegraphics[width=8.4cm]{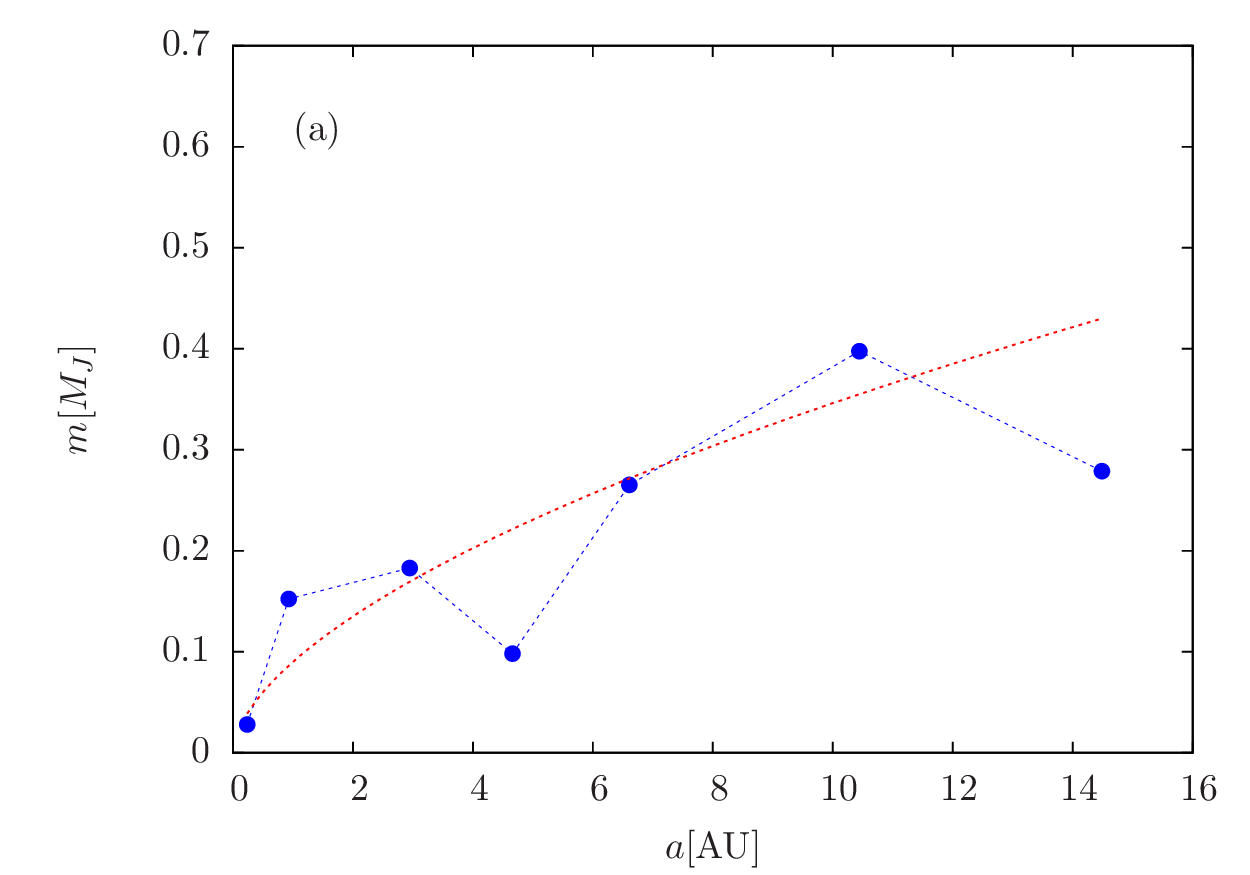}\\
  \includegraphics[width=8.4cm]{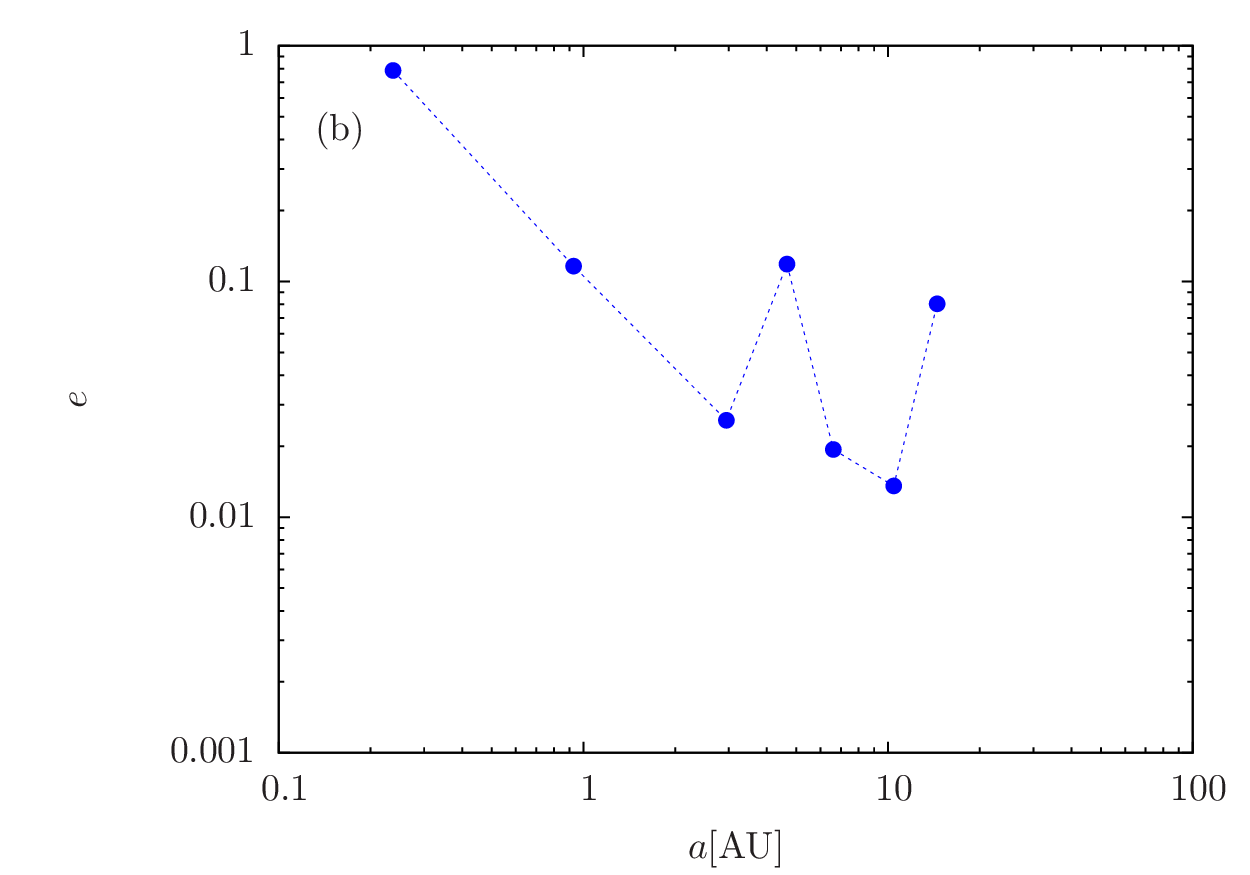}
  \caption{\label{fig9}%
    Same as Fig.~\ref{fig8} for a randomly selected planetary system
    obtained in the oligarchic growth model. The continuous curve in
    the top panel is the power-law fit of the averaged data displayed
    in Fig.~\ref{fig8}(a).}
\end{figure}

The results of {\mybf these} simulations are shown in Fig.~\ref{fig6}, for the
combined results of the ensemble, and in Fig.~\ref{fig7} for an
individual planetary system taken at random. As shown in the results
for the ensemble, the mass increases with the semi-major axis up to
$a\approx 12.6$~AU, where the feature associated with the location of
the {\mybf outermost} planet appears. Fitting
Eq.~\ref{eq:massfit} with the local average mass up to $9.34$~AU
yields the power-law exponent $\nu\approx 0.74$. This value of $\nu$
is even larger than the one obtained for runaway accretion. The
mass-scale is larger as well, being enhanced by about $25\%$ with
respect to the mass-selection mechanism; the maximum mass reaches
approximately $0.53~M_J$.  Correspondingly, the average final number
of planets is further reduced to $9.3\pm0.7$. As done before, the
limit for fitting the power-law was considered due to the appearance
of the strong oscillations in the averaged data, as illustrated in
Fig.~\ref{fig6}(a).

The density profile displayed in Fig.~\ref{fig6}(a) resembles the one
obtained for the orderly growth model, being more uniform than the
density profile for runaway growth. Yet, there is a qualitative
difference with respect to the former results, which is manifested by
a gap in the mass--semi-major axis diagram. Indeed, there is a region
in the $m$ vs $a$ diagram, defined beyond $10.7$~AU (for small masses), 
where no planet is formed. That is, planets formed beyond $\sim10$~AU
have a mass above a certain lower bound, which depends on
$a$. Actually, a similar feature can also be observed for small values
of $a$. We also note the less-massive planets formed at intermediate
semi-major axis are consistently more massive than the corresponding
ones for the mass-selection mechanism.

With respect to the eccentricities of the ensemble, as illustrated in
Fig.~\ref{fig6}(b), we confirm the overall resemblance with the
results obtained for orderly growth, where $\theta$ is a
uniformly-distributed random variable in the proper interval. In the
present case, the distribution of final planetary eccentricities
yields the possibility of larger values of the eccentricity. This can
be noticed, e.g., in the branch of small $a$ associated with the
location of the innermost planets, and for large values in the sudden
increase of the average eccentricity of the outermost planets.

\subsection{Mass- and velocity-selection mechanism:
 oligarchic growth}
\label{variation3}

We consider now the combined effect of the mass- and
velocity-selection mechanisms. To this end, we select a particle
according to the mass-selection mechanism described previously, and
then select the relative orientation $\theta$ of the elliptic orbits
of the colliding particles such that $v_{\rm rel}$ is a minimum. This
case is somewhat analogous to {\it oligarchic growth}
models~\citep{Kokubo1998}, where after an initial runaway phase,
gravitational focussing is enhanced by all the dominating protoplanets
but on a slower time-scale. We shall thus use this name to refer to
the present case. The combined results of $4096$ simulations are shown
in Fig.~\ref{fig8}, and the result of one randomly selected
realisation in Fig.~\ref{fig9}.

As illustrated in Fig.~\ref{fig8}(a), the results display an
increasing behaviour of the mass upon the semi-major axis, until the
feature associated with the position of the outermost planet
appears. In this case, fitting the averaged data with
Eq.~\ref{eq:massfit} in the interval $a\in[0,8.6]$ yields $\nu\approx
0.58$. This value is comparable (marginally smaller) than the value
obtained for orderly growth. Despite this, the overall mass-scale is
slightly larger than in previous cases, reaching the value $0.63~M_J$,
consequently yielding a smaller average number of formed planets:
$6.9\pm0.7$.

With regards to the final eccentricities, Fig.~\ref{fig8}(b)
illustrates the enhancement of the final eccentricity of the
orbits. Indeed, we observe an increased variance of the final
eccentricity at all scales of the semi-major axis, as well as their
average value.  In particular, we note that for the outermost planet,
the corresponding eccentricity may reach values close to
$0.2$. Interestingly, this indicates that the distribution of $C_{\rm
  final}$ has larger values than in the other cases.

\begin{figure}
  \includegraphics[width=8.4cm]{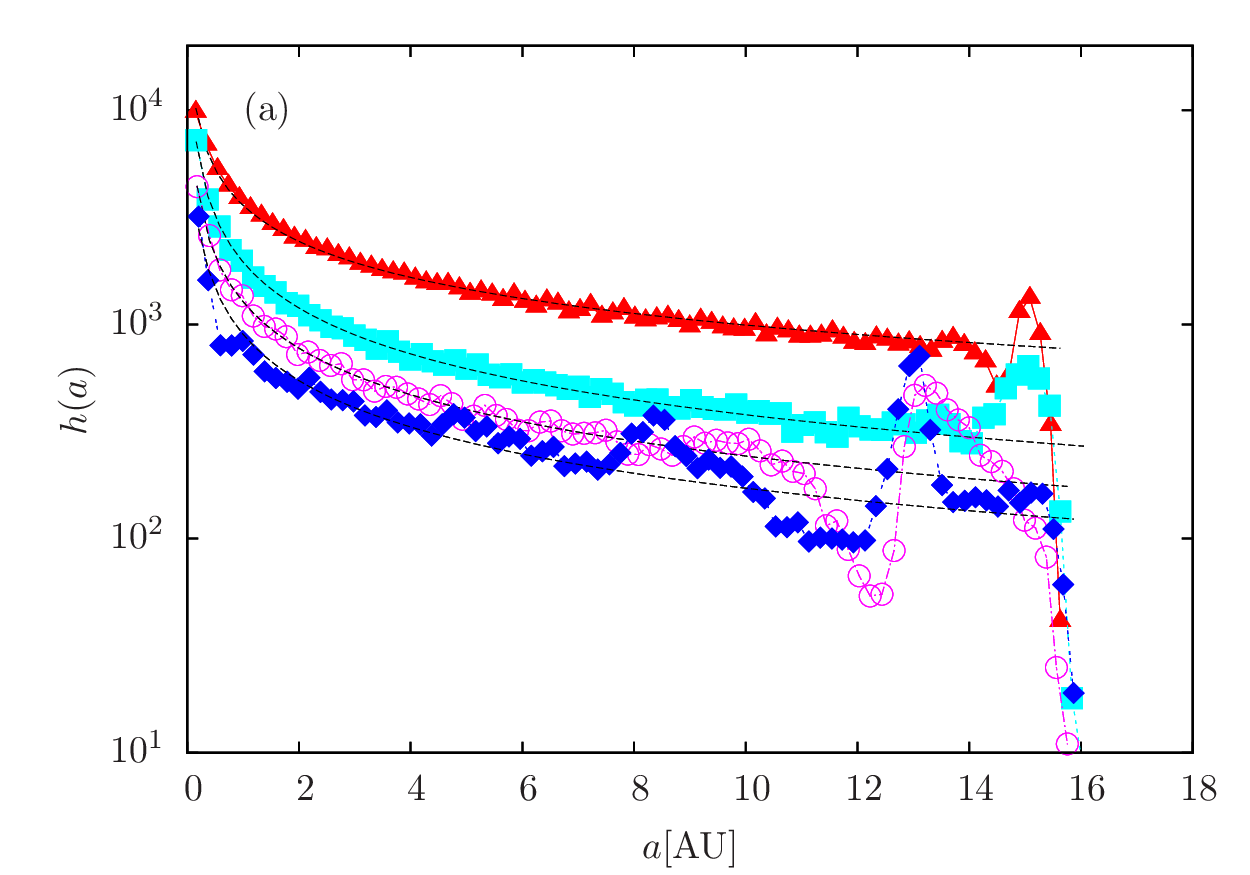}\\
  \includegraphics[width=8.4cm]{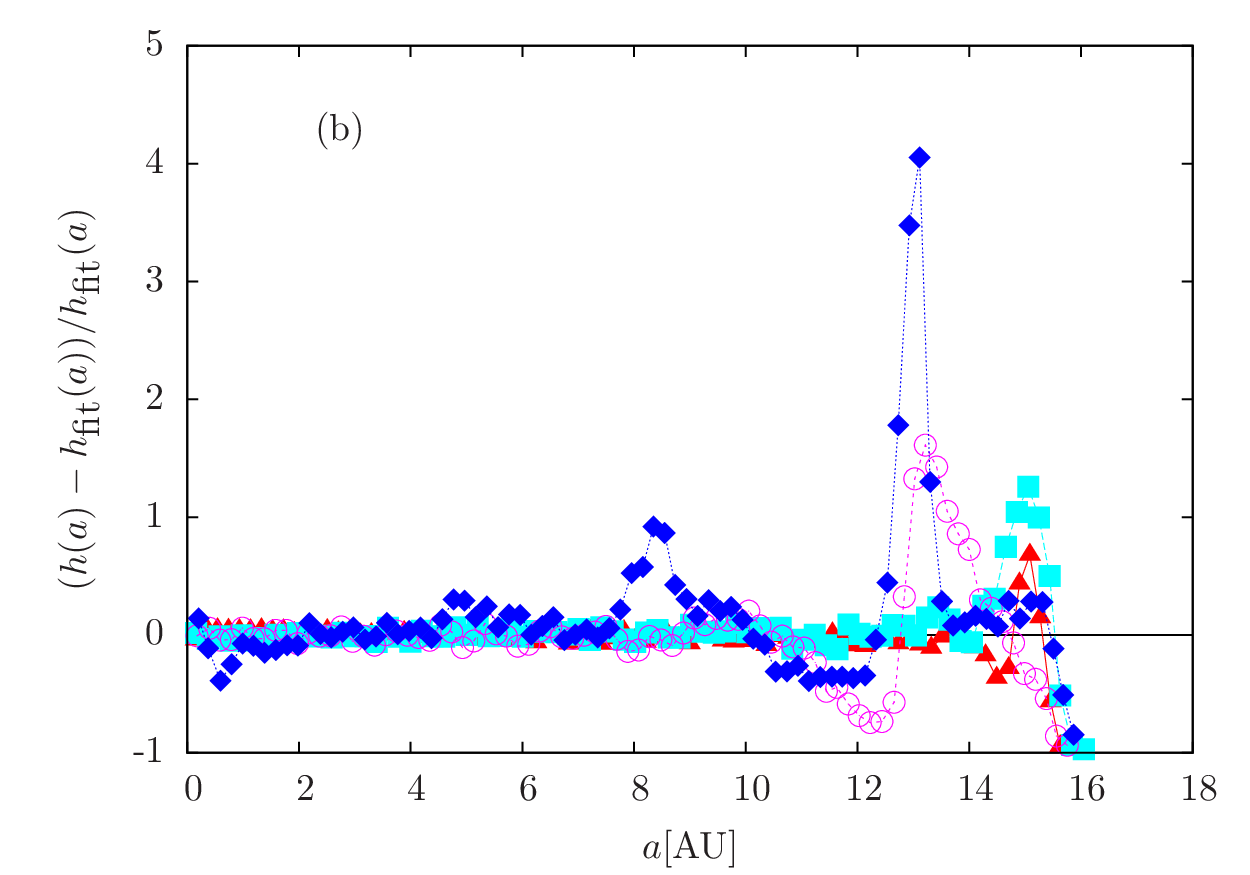}
  \caption{\label{fig10}%
    (a)~Number of planets $h(a)$ formed in an interval of size $\Delta
    a\sim 0.2$~AU around $a$. The symbols identify the different
    growth models: Triangles (red) for orderly growth (Laskar's model
    with conservation of $L_{\rm tot}$), squares (cyan) for runaway
    growth, empty circles (magenta) for the velocity-selection
    mechanism, and filled circles (blue) for oligarchic growth. The
    dashed lines are the best-fitting power-law $h_{\rm fit}(a)$ for
    each case. (b)~Relative residuals with respect to $h_{\rm
      fit}(a)$. The peaks indicate a systematic enhancement of the
    formation of planets in certain specific locations.}
\end{figure}

The density profiles displayed in the mass and eccentricity diagrams
are interesting and differ clearly in the structure from the other
cases. Indeed, not only the location of the outermost and innermost
planets is noticeable, but actually the location of other planets can
be estimated from the darken clumps that appear in the figures. This
can also be observed in the oscillations displayed by the averaged
data, which, in comparison with the cases discussed before, are more
prominent and appear even for smaller values of $a$.

We illustrate these observations in Fig.~\ref{fig10}. In
Fig.~\ref{fig10}(a) we plot the number of final planets $h(a)$ (on a
log scale) located in an interval of size $\Delta a\sim 0.2$~AU
centred around a semi-major axis $a$, for the four different growth
models discussed. The dashed curves are the best-fitting power-law 
corresponding to each data set, which we denote $h_{\rm fit}(a)$. In
all growth models, we observe a clear excess of particles (with
respect to the best-fitting) towards the outer edge of the planetary
systems, which is associated with the location of the outermost
planet. The excess in the particle number with respect to $h_{\rm
  fit}(a)$ is close to or above $600$ particles in the orderly ($a\sim
15.1$~AU) and oligarchic ($a\sim 13.06$~AU) models, being more
moderate in the other cases. We observe other peaks in the oligarchic
growth model, at $a\sim 8.35$~AU and at $a\sim 4.8$~AU; in each case
the excess of particles with respect to $h_{\rm fit}(a)$ becomes more
moderate as the semi-major axis decreases. These observations point
out the existence of other specific locations where planets are formed
preferentially in the oligarchic model, once the AMD stability
criterion is satisfied.

In order to show that such locations are indeed a systematic property
with full statistical meaning, in Fig.~\ref{fig10}(b) we display the
relative residuals $[h(a)-h_{\rm fit}(a)]/h_{\rm fit}(a)$ of the
corresponding fit for the four growth models. The location of the
outermost peak in all models is at least a $65\%$ off the
best-fitting. Notice that the case of the oligarchic growth
corresponds to a deviation over $\sim 400\%$ from the
best-fitting. With regards to the secondary peaks observed in the
oligarchic growth model, they are $\sim90\%$ and $\sim 29\%$ in excess
with respect to $h_{\rm fit}(a)$, in decreasing order of $a$,
respectively. This shows that these peaks are due to systematic
correlations in the data, i.e., a systematic formation of planets in
specific locations in this oligarchic growth model; in other words, an
enhancement of the probability to find planets at those specific
locations.

Figure~\ref{fig10}(b) shows, in addition, that the peak associated
with the outermost planet, that lies at the smallest semi-major axis,
corresponds to the oligarchic growth one. This shows that migration is
promoted by the combined effect of mass and velocity-selection
mechanisms, as in the oligarchic growth model. Taking as reference the
orderly growth model and comparing the results for mass and
velocity-selection mechanisms, we see that the effect is largely
dominated by the the latter mechanism.

\section{Universality}
\label{univ}

An important question in relation to the results presented in the
previous section, is what happens if the important physical parameters
$L_{\rm tot}$, $M_{\rm disc}$, and $M_0$ have different values. %
Put differently, how do the results depend upon the
specific values of $L_{\rm tot}$, $M_{\rm disc}$ and $M_0$. 
Notice that different values of these physical parameters may affect 
the radial extension of the disc, cf. Eq.~\ref{eq:lztot}. In this
section, we address this question and show that the results are
universal. Here, by universality, we mean that the results of
different parameters are statistically the same under properly defined
scaled variables. Notice that this universality may not hold by changing other
parameters of the model such as $e_{\rm max}$, $\rho_a(a)$ or
$\beta$~\citep{MenaPhD}. This question is clearly of central
importance for a comparison with the observations of exo-solar
planetary systems.

To this end, we shall define the length-scale
\begin{equation}
\label{Rscale}
  R_{\rm scale} = [L_{\rm tot}/M_{\rm disc}]^2/\mu,
\end{equation}
which is defined only in terms of the dynamically invariant quantities
of the underlying many-body Hamiltonian
(Eqs.~\ref{eq:hamiltonian0}--\ref{eq:hamiltonian1}). We emphasize here
that $\mu=G M_0$ includes the dependence on the mass of the host star
$M_0$, and is not fixed to $4 \pi^2$ unless $M_0=M_\odot$. The
dimensionless scaled variables are thus given by
\begin{eqnarray}
\label{eq:ScaledDist}
a^\prime &=& a/R_{\rm scale}, \\
\label{eq:ScaledMass}
  m^\prime &=& m/M_{\rm disc}.
\end{eqnarray}
According to these equations, the masses of the planets formed scale
linearly with the mass of the disc, i.e., $M_{\rm disc}$ defines
simply the overall mass scale of the simulations. In turn, the
semi-major axes scale linearly with $M_0$, and quadratically with the
ratio of the total angular momentum and the mass of the disc.

We compare now equivalent calculations for the same growth model 
using different physical parameters. We shall use the oligarchic growth
model of Section~\ref{variation3} as example, and compare with
calculations performed for the same values of $M_0$ and $M_{\rm
  disc}$, which therefore will not change the mass-scale of the formed
planets, but with the total angular momentum of the system fixed to
$L^\prime_{\rm tot}=3.479\times 10^{-2} M_\odot \, {\rm AU}^2 \, {\rm
  yr}^{-1}$. The latter value is $\sim 1.57$ times the value used in
Section~\ref{variation3}.

\begin{figure}
  \includegraphics[width=8.4cm]{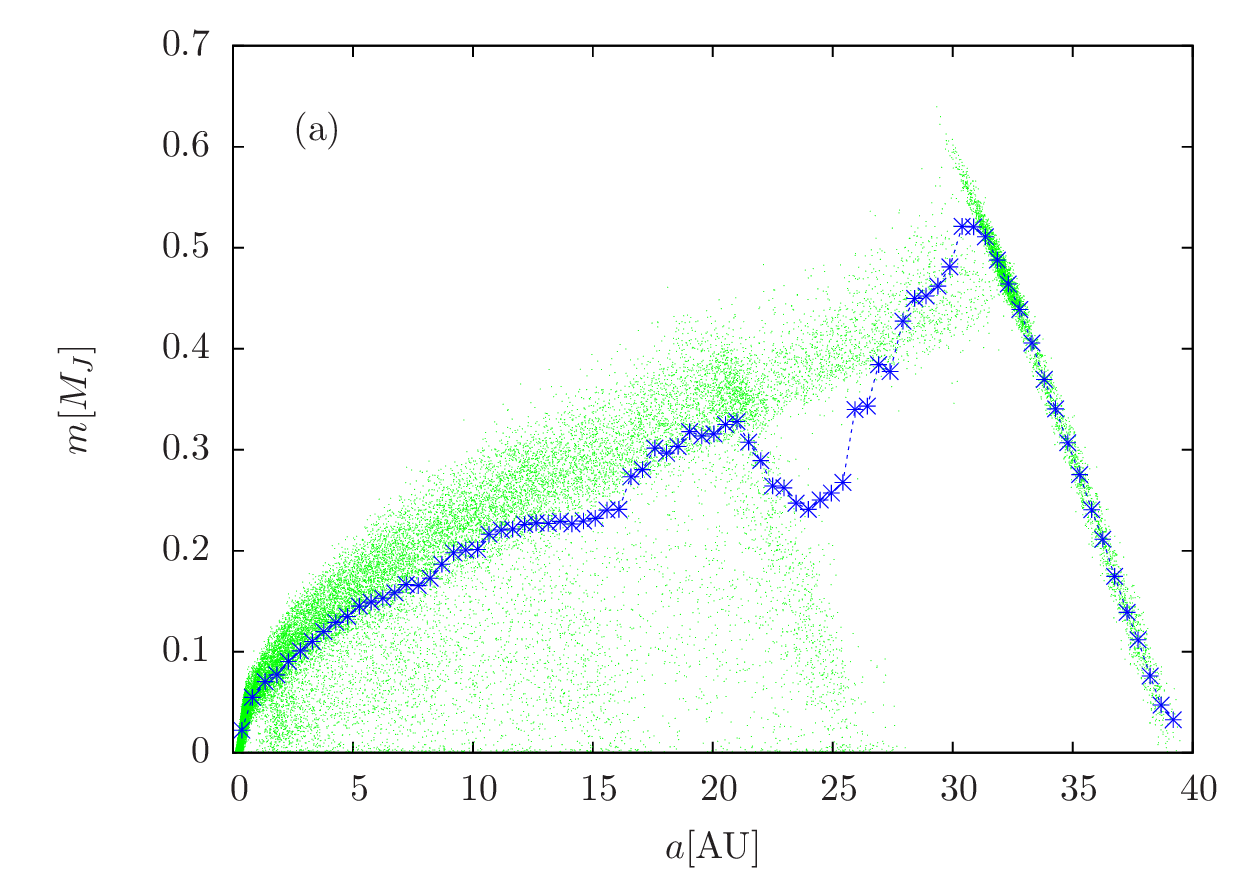}\\
  \includegraphics[width=8.4cm]{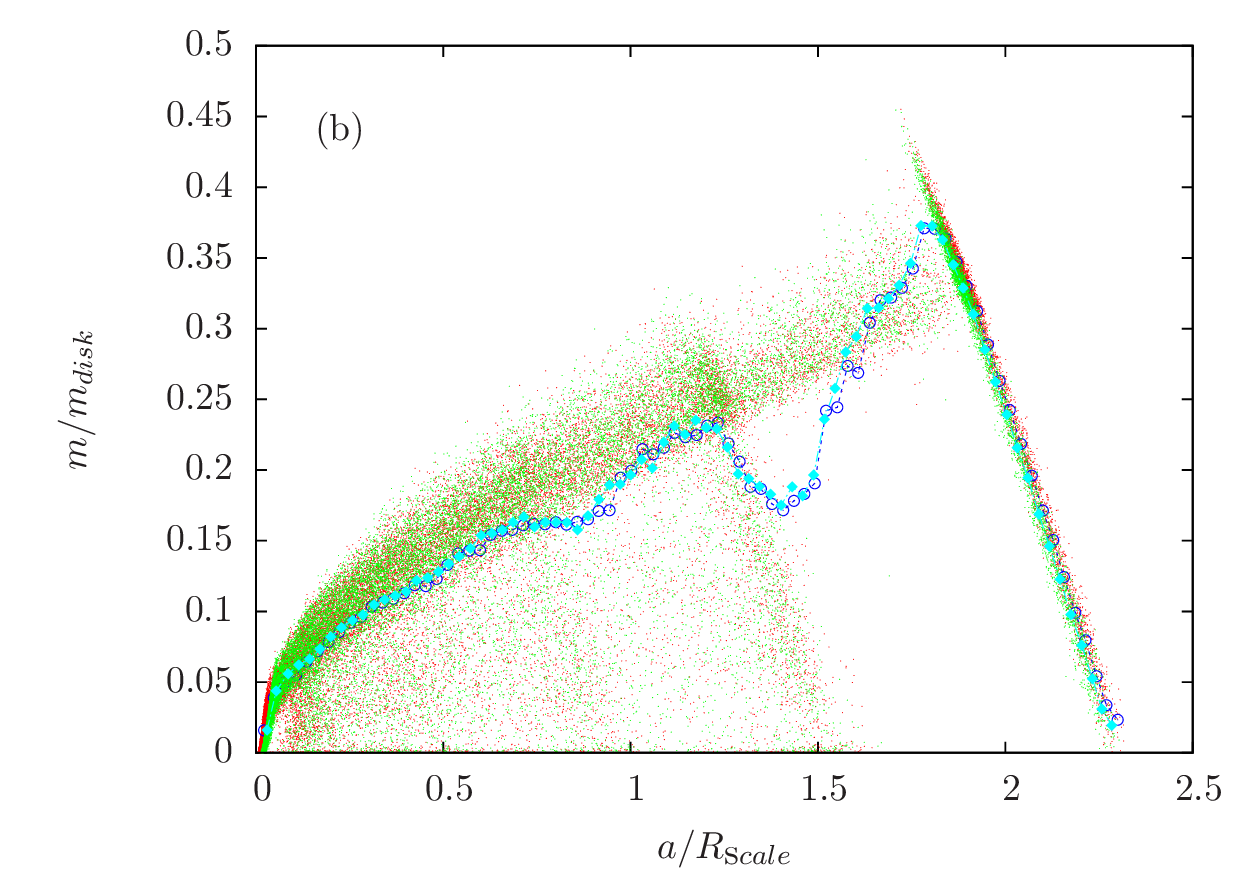}%
  \caption{\label{fig11}%
    (a)~Mass--semi-major axis diagram for the combined results of
    $4096$ simulations of the oligarchic growth model with total
    angular momentum of the system set to $L^\prime_{\rm tot}=3.479
    \times 10^{-2} M_\odot \, {\rm AU}^2 \, {\rm yr}^{-1}$ ($M_{\rm
      disc}$ and $M_0$ are the same as in the simulations of
    Section~\ref{variation3}). (b)~Same as the top panel but using the
    scaled variables, Eq.~\ref{eq:ScaledDist} and~\ref{eq:ScaledMass},
    displaying the data used in (a) and the data used in
    Fig.~\ref{fig8}(a).}
\end{figure}

Figure~\ref{fig11}(a) shows the mass--semi-major axis diagram for this
case. The figure is quite similar, in a qualitative sense, to
Fig.~\ref{fig8}(a). Quantitatively, the vertical scales of both
figures coincide, as expected. The horizontal scale is enlarged by a
factor close to $2.5$ (actually, $(L^\prime_{\rm tot}/L_{\rm
  tot})^2$). In Fig.~\ref{fig11}(b) we display the same correlation
diagram using the scaled variables defined in Eqs.~\ref{eq:ScaledDist}
and~\ref{eq:ScaledMass}, for the data of both parameter sets. The
difficulty to distinguish the data from one case to the other clearly
illustrates the meaning of universality. We emphasize that the same 
scaling holds for the position of the peaks of the relative residuals that 
mark the systematic formation of planets at certain specific locations.
Equivalent results can be
obtained by varying the values of $M_{\rm disc}$ and $M_0$.

The universal property illustrated in Fig.~\ref{fig11} is a
consequence of the fact that the ($N+1$)-body Hamiltonian
(Eqs.~\ref{eq:hamiltonian0}--\ref{eq:hamiltonian1}) includes only
terms which build up a homogeneous potential (of degree -1), thus
being invariant under appropriate
scaling~\citep{LandauMechanics}. Therefore, universality actually
holds for any dynamical variable and also for all the growth models
discussed here. This property opens another perspective to
statistically analyse the data of exo-solar planetary systems.

Yet, the scaling property of universality may break down if
other parameters used in the model strongly influence the results of
the simulations. Changing the functional form of $\rho_a(a)$ will
change the scaling properties of the mass in terms of the semi-major
axis, which can be inferred from the analytical results
of~{\mybf\citet{Laskar2000}}, cf. Eq.~\ref{eq:mass}.  Another more subtle
situation arises, for example, when a much larger value of
$L^\prime_{\rm tot}$ is considered, keeping all other parameters
fixed. In this case, the initial planetesimals disc will be more
extended with respect to the parameter corresponding value $a_{\rm
  max}$, with the same functional form of the initial $\rho_a(a)$, thus
making up a much fainter planetesimal disc due to normalization. If
the density is not taken consistently, the value of $e_{\rm max}$ used
may become important in the simulations, since some collisions of
particles may not take place for $L^\prime_{\rm tot}$. Adjusting the
number of particles at the beginning of the simulations yields
universality again.

\section{Summary and concluding remarks}
\label{concl}

In this paper, we have studied some physically-motivated variations on
Laskar's simple model of accretion and 
evolution~{\mybf\citep{Laskar2000,LaskarPrep}}, simulating the formation 
and dynamics of planetary systems in the secular
limit, thus excluding effects of mean-motion resonances. Fulfilling
the AMD stability criterion ensures that planetary collisions are no
longer possible for the averaged system, but might be possible if
mean-motion resonances were included. The variations we incorporated
include important situations which are relevant to the formation of
planetary systems according to the current
understanding~\citep{Safronov,Pollack96,Armitage:2007arXiv}. In
particular, we implemented the model with a strict conservation of
the total angular momentum of the system during the simulations. We
also addressed the implications that arise from including in the
accretion probability distribution a dependence on the mass of the
planetesimals, and on the relative velocity of the colliding particles
at the collision point. These are important factors that influence the
accretion rate processes. In our implementation of these physical
effects we have tried to maintain the simplicity of Laskar's original
model. An important limitation of our models, inherited from Laskar's 
simplified model, is the lack of a true dynamical evolution, which thus 
prevents us from answering questions involving or addressing the 
physical time in the formation processes.

Our statistical analysis was based on the mass--semi-major axis
correlations of the formed systems, using the current values of the
total angular momentum and mass of the planets of the Solar System as
main parameters, though we included also remarks on the final
eccentricities. For the comparisons, we consider as reference Laskar's
model imposing the conservation of the total angular momentum of the
system, which we referred to as orderly growth. This case shows a
power-law behaviour of the mass of the formed planets in terms of the
semi-major axes in accordance to Laskar's prediction, but with a
different value of the exponent. At the edges of the disc, our
statistical results showed distinctive features deviating from this
power-law behaviour, which are associated with the location of the
inner- and outer-most planets. Introducing a mass-selection mechanisms
where more massive bodies have an enhanced probability to accrete,
thus modelling a kin of runaway growth, we found a comparatively
smaller number of planets formed, which are more
massive. Interestingly, we observed a systematic occurrence of rather
small-mass remnants embedded in the planetary system. We also
considered a velocity-selection mechanism by only allowing accretion
when Safronov's condition is satisfied; in this case, our results
showed that the AMD stability criterion was not fulfilled in
general. We thus relaxed this condition, and fixed the relative
orientation of the elliptic orbits of the colliding bodies to the
value corresponding to the minimum of the relative velocity. This
criterion selects the quasi-tangential collisions, which take longer
times of interaction, making physically plausible accretion. In this
case, we found that the average number of planets was further reduced,
which are then more massive, at the end of the
simulations. Furthermore, the results manifest the appearance of a gap
in the mass--semi-major axis diagram, which defines a lower bound for
the masses of the planets at either edge, that depends on their
location.

We also considered the combination of the mass- and velocity-selection
mechanisms, in what we called the oligarchic growth model. In this
case, the number of final planets was further reduced and their masses
increased, with the peak reaching values over $0.6 M_J$. An
interesting result is the systematic appearance of planets at specific
locations, manifested as a definite excess of planets formed in this
growth model. This property is manifested as a clustering of planets,
i.e., an enhancement in the density profile in definite regions of the
mass--semi-major axis diagram. We emphasize that, despite the
stochastic nature of the model, the combined mass- and
velocity-selection mechanisms induce strong correlations that yield
this clustering. We observed from the location of the outermost
planet, that the strongest migration occurs precisely in this
oligarchic growth, due fundamentally to the velocity-selection
mechanism, i.e., gravitational focusing.

We also addressed the question of variations to the main physical
parameters of the model. We found that the results are statistically
invariant with respect to such changes, whenever they are expressed in
terms of appropriate scaled variables. The masses scale linearly with
the mass of the disc, and the semi-major axes scale according to
Eqs.~\ref{Rscale} and~\ref{eq:ScaledDist}, involving the angular
momentum, the mass of the disc and the mass of the central star. In
this sense, the results are universal. Universality holds in
particular when the same functional form of the initial distributions
$\rho_a(a)$ and $\rho_e(e)$ are maintained. This property follows from 
the mechanical similarity of the Hamiltonian~\citep{LandauMechanics}.

In most of our simulations, we used the parameters describing the
current state of the Solar System, and an initial homogeneous linear
mass-density distribution $\rho_a(a)$. 
We expect that changing the initial linear mass-density distribution 
influences the power-law behaviour obtained, essentially as it does in 
the analytical results of {\mybf\citet{Laskar2000}}. Comparing the masses 
of the planets formed in the simulations with those of the Solar System, in
our simulations the inner planets are more massive, and the outer
planets are thus not massive enough. Yet, it is interesting to remark
that the occurrence of the specific locations (semi-major axes) where
the planets form, obtained for the oligarchic growth model, are
consistent with the initial conditions used for three (out of four) of
the major planets considered in the simulations of the architecture of
the Solar System performed within the first
Nice model by~\citet{Tsiganis2005};
as mentioned, this unexpected agreement does not hold for the
masses~\citep{Thomes2003}. Considering that our results are
qualitative, such a partial quantitative agreement is encouraging to
construct more realistic models, that fully explain the initial
conditions of the Nice model in a statistical robust sense.

Future work along these lines will be aimed to incorporate in the
model effects related to mean-motion resonances~\citep{Malhotra95},
which are known to be important in the architecture of the Solar
System~\citep{Tsiganis2005,Morbidelli2007,Morbidelli2009}. 
Furthermore, improving on the assumed
form of the initial linear mass-density distribution $\rho_a(a)$ may
yield better comparison of the mass distributions. For the concrete
case of the Solar System, \citet{Laskar1997,Laskar2000} suggested that
a better modelling may be an initial linear mass-density distribution
that considers separately the inner and outer solar systems. Other
possibilities include the distinction of particles as gas or heavy elements,
or the addition of an outer planetesimal disc, as assumed
by the Nice model.

\section*{Acknowledgments}
{\mybf It is our pleasure to thank Jacques Laskar for providing us with a 
copy of his work prior to publication \citep{LaskarPrep}, for discussions 
and hospitality. We are also thankful to Kartik Kumar, Hern\'an
Larralde and Fr\`ederic Masset for helpful discussions and suggestions}. We
would like to acknowledge financial support from the projects
IN-110110 (DGAPA-UNAM) and 79988 (CONACyT).

\bsp

\label{lastpage}

\end{document}